 \documentclass[showpacs,reprint,aps,pre]{revtex4-2}
 \usepackage{amsfonts}
 \usepackage{natbib}
 \usepackage{amsmath}
 \usepackage{amssymb}
\usepackage[colorlinks,citecolor=blue,linkcolor=blue]{hyperref} 
 \usepackage{mathtools}
 \usepackage[caption=false]{subfig}
 
 \DeclarePairedDelimiter\floor{\lfloor}{\rfloor}
 \DeclareMathOperator{\arccosh}{arccosh}
\DeclareMathOperator{\arcsinh}{arcsinh}
 
\begin{document}
  \title{Chaos and quantization of the three-particle generic
  Fermi-Pasta-Ulam-Tsingou model I: Density of states and spectral statistics}

  \author{Hua Yan}
  \email{yanhua@ustc.edu.cn}
  \affiliation{CAMTP - Center for Applied Mathematics and Theoretical
  Physics, University of Maribor, Mladinska 3, SI-2000 Maribor, Slovenia,
  European Union}
  \author{Marko Robnik}
  \email{Robnik@uni-mb.si}
  \affiliation{CAMTP - Center for Applied Mathematics and Theoretical
  Physics, University of Maribor, Mladinska 3, SI-2000 Maribor, Slovenia,
  European Union}

  \date{\today}

  \begin{abstract}
    We study the mixed-type classical dynamics of the three-particle
    Fermi-Pasta-Ulam-Tsingou (FPUT) model in relationship with its quantum counterpart,
    and present new results on aspects of quantum chaos in this system.
    First we derive for the general N-particle FPUT system the
    transformation to the normal mode representation. Then we specialize
    to the three-particle FPUT case, and derive analytically
    the semiclassical energy density of states, and its derivatives in
    which different singularies are determined, using the Thomas-Fermi rule.
    The result perfectly agrees with the numerical energy density from
    the Krylov subspace method,
    as well as with the energy density obtained by the method of
    quantum typicality. Here, in paper I, we concentrate on the energy
    level statistics (level spacing and spacing ratios),
    in all classical dynamical regimes of interest: the almost entirely
    regular, the entirely chaotic, and the mixed-type regimes. We clearly confirm,
    correspondingly, the Poissonian statistics, the GOE statistics, and the
    Berry-Robnik-Brody (BRB) statistics in the mixed-type regime. It is found
    that the BRB level spacing distribution perfectly fits the numerical data.
    The extracted quantum Berry-Robnik parameter is found to agree with the classical
    value within better than one percent. We discuss the role of localization of chaotic
    eigenstates, and its appearances, in relation to the classical phase space
    structure (Poincar\'e and SALI plots), whose details will be presented in
    paper II, where the structure and the statistical properties of the
    Husimi functions in the quantum phase space will be studied.
    \end{abstract}
  
  
  \maketitle

  \section{Introduction}
  \label{sec1}
  
  The Fermi-Pasta-Ulam-Tsingou (FPUT) model \cite{fermi1955studies,dauxois2008fermi}, is the beginning of the massive use of numerical methods implemented on electronic computers in the physical sciences \cite{lepri20231953},  has a long history and still, even after 68 years, presents intriguing open questions due
  to its rich classical, and quantal, structures an dynamics, not only in the case of large N-body system, but even in the simplest nontrivial case of just three
  particles (N=3). The study of this system certainly initiated new developments in fundamental physics by the pioneers in classical and quantum chaos, and in statistical physics in general. There are hundreds of important papers in the literature dealing with FPUT, which cannot be listed and discussed here. We just refer
  to some important reviews \cite{arnol1963small,arnold2009proof,moser2001stable,livi1985equipartition,ford1992fermi,berman2005fermi,gallavotti2007fermi,onorato2015route,gallone2022burgers} covering the history of research starting from the initial paradox to its fundamental resolution and the understanding of the general behavior of the model,
  which turned out to be of generic, mixed type, thereby giving rise to insights in both scenarios, the  integrable and the chaotic one. 
  
  Our aim in this paper is to perform a detailed study of the classical dynamics of the system with three particles (N=3), and of its quantum counterpart. We perform the detailed analysis of the aspects of quantum chaos, by
  looking at the energy spectra and their statistical properties, followed (in paper II) by the analysis of the Husimi functions of the eigenstates. After introducing the normal modes the model is reduced to the  motion of just one particle in two-dimensional potential. The cubic
  case ($\alpha$-FPUT) is the celebrated Henon-Heiles Hamiltonian \cite{henon1964applicability}, which is nonintegrable and of the mixed type, while the pure quartic potential, $\beta$-FPUT, is integrable, as the angular momentum is preserved. We treat also the general case, $\alpha \beta$-FPUT, which also is of the mixed type, but becomes increasingly more regular as the parameter $\beta \rightarrow\infty$, so that the quartic potential asymptotically dominates the dynamics.
  
  Our main results comprise a detailed analysis of the classical dynamics, also using the SALI chaos detecting technique, needed for the study of quantum chaos. We then perform the exact analytic calculation of the semiclassical density of states (DOS) using the Thomas-Fermi rule, needed for the unfolding of the quantum energy spectra. The result agrees perfectly with the estimate of DOS using the method of quantum typicality, as well as with the numerical DOS. The level statistics clearly displays transition from the predominantly  integrable regime to the chaotic regime, as exhibited by the level spacing distributions and the mean spacing ratio. As a side product of this study, we also found
  a clear functional relationship between the normalized mean spacing ratio and the Berry-Robnik parameter ($\mu_c$, which is the relative size of the classically
  chaotic region in the phase space). By applying the Berry-Robnik-Brody level spacing distribution (BRB), we find perfect agreement between the quantum and classical value of $\mu_c$. As extracted from BRB, we
  find that most regimes are such that the chaotic eigenstates are largely extended, or just weakly localized. The detailed study of the structure
  and statistics of the Husimi functions, as well as the power-law decay of the fraction of the mixed eigenstates will be treated in paper II.
  
  The paper is structured as follows. In Sec. \ref{sec2} we perform the reduction of the general FPUT system to normal modes.  In Sec. \ref{sec3} we specialize to the three particle FPUT system, and  analyze in detail the transition from regularity to chaos in the classical dynamics by means of the Poincar\'e surfaces of section (SOS) as well as by SALI plots, and estimate the relative size of the chaotic component $\mu_c$. In Sec. \ref{sec4} we perform the quantization of the three particle FPUT system by introducing the rotated bosonic operators, leading to the matrix elements in the circular two-mode basis. We discuss the role of symmetries in Sec. \ref{sec4.3}. In Sec. \ref{sec5} we calculate DOS by use of the Thomas-Fermi rule, the quantum typicality method, and by means of the numerical energy spectra. They all agree very well. In Sec. \ref{sec6} we analyze the statistical properties of energy spectra. In Sec. \ref{sec7} we conclude and discuss the outlook.

  \section{Introducing the normal modes for the general FPUT system}
  \label{sec2}
  We consider a chain of $N$ moving particles with nearest neighbor interaction given by a potential $V$,  the Hamiltonian of such a system is given by
  \begin{align}
    \label{eq:fpu-pbc}
    H=\sum_{j=1}^N\left(\frac{y_j^2}{2} + V(x_{j+1}-x_j)\right),
  \end{align}
  with the periodic boundary condition (PBC) $x_1=x_{N+1}$, where $x_j$ are the displacement of the particles with respect to the equilibrium positions and $y_j$ are the corresponding momenta. It should be noted that the original Fermi-Pasta-Ulam-Tsingou (FPUT) report used fixed boundary condition (FBC) $x_1=x_{N+1}=0$, having the following Hamiltonian
  \begin{align}
    H=\sum_{j=2}^{N}\frac{y_j^2}{2} + \sum_{i=1}^{N}V(x_{j+1}-x_j).
  \end{align}
  The  $\alpha\beta$-FPUT model is given by the following choice of the potential
  \begin{align}
    V(s)=\frac{1}{2}s^2+\frac{\alpha}{3}s^3+\frac{\beta}{4}s^4,
  \end{align}
  where we will indicate as $\alpha$-FPUT for the case $\beta=0$, and as $\beta$-FPUT the case $\alpha=0$. 
  
  The equations of motion from Eq. \eqref{eq:fpu-pbc} with the quadratic form of potential ($\alpha=\beta=0$) are
  \begin{align}
    \ddot{x_j}=-\frac{\partial H}{\partial x_j}= x_{j+1}-2x_j +x_{j-1}, \quad x_j=x_{j+N}.
  \end{align}
  Trying an oscillatory solution of the form $x_j=a_j e^{-i\omega t}$ and substituting this trial solution into the equations of motion, one gets the equations of amplitude factors
  \begin{align}
    (\omega^2-2)a_j+a_{j+1}+a_{j-1}=0, \quad a_j=a_{N+j}.
  \end{align}
  These $N$ linear homogeneous equations $A_pa=0$ for $a=[a_1,\cdots,a_N]^T$ have a nontrivial solution only if the determinant of the coefficient matrix $A_p$ vanishes, where $A_p$ can be written explicitly as
  \begin{equation}
    A_p=\begin{bmatrix}
    \omega^2-2 & 1 &0& \dots &1  \\
    1 & \omega^2-2 &1&  \dots&0 \\
    0 & 1 & \omega^2-2&\dots  & 0\\
    \vdots&\vdots & \vdots &\vdots &\vdots\\
    1& 0&\dots& 1&\omega^2-2 \\
    \end{bmatrix},
  \end{equation}
 is a Toeplitz matrix \cite{horn2012matrix}, for which we know the analytical form of all the eigenvalues, which yields
  \begin{align}
    \label{eq:norm-pbc}
    \omega_k = 2\sin\frac{\pi k}{N}, \quad k=0,\dots ,N-1.
  \end{align}
  $\omega_k$ are also the well-known normal mode frequencies. There 
  is a zero normal mode in the PBC case, obviously due to the translational symmetry. From the diagonalization of the Toeplitz matrix $A_p$, the normal modes can be constructed from the eigenvectors. For $N$ odd and $k <N/2$, $(Q_k,P_k)$ the new coordinates and momenta of the harmonic normal modes are
  \begin{align}
      \label{eq:normal-modes}
      &Q_0=\frac{1}{\sqrt{N}}\sum_{j=1}^Nx_j,\quad 
      Q_k=\sqrt{\frac{2}{N}}\sum_{j=1}^N x_j\sin\frac{2jk\pi}{N},\nonumber \\
      &Q_{N-k}=\sqrt{\frac{2}{N}}\sum_{j=1}^N x_j\cos\frac{2jk\pi}{N}.
  \end{align}
  Specifically, if $N$ is even, we have additionally
  \begin{align}
      Q_{N/2}=\frac{1}{\sqrt{N}}\sum_{j=1}^N(-1)^jx_j,
  \end{align}
 and similar definitions for $P_k$. The resulting energy of $k$-th normal mode is  $ \epsilon_k = \frac{1}{2}(P_k^2+\omega_k^2Q_k^2)$. Applying the inverse (Fourier) transform, one has the transformation from $Q\to x$ (similar transformation for $P\to y$) as
  \begin{align}
      \label{eq:inverse-odd}
      x_j=\frac{Q_0}{\sqrt{N}}+\sqrt{\frac{2}{N}}\sum_{k=1}^{\floor{N/2}}\big(Q_k\sin\frac{2jk\pi}{N} + Q_{N-k}\cos\frac{2jk\pi}{N}\big),
  \end{align}
for odd $N$. If $N$ is even, we have
\begin{widetext}
  \begin{equation}
    \label{eq:inverse-even}
    x_j=\frac{1}{\sqrt{N}}Q_0+\frac{(-1)^j}{\sqrt{N}}Q_{N/2}+\sqrt{\frac{2}{N}}\sum_{k=1}^{N/2-1}\big(Q_k\sin\frac{2jk\pi}{N}+ Q_{N-k}\cos\frac{2jk\pi}{N}\big).
\end{equation}
\end{widetext}

  \section{Emergence and analysis of classical chaos}
  \label{sec3}

  \begin{figure*}
    \centering
    \subfloat{{\includegraphics[width=0.8\textwidth]{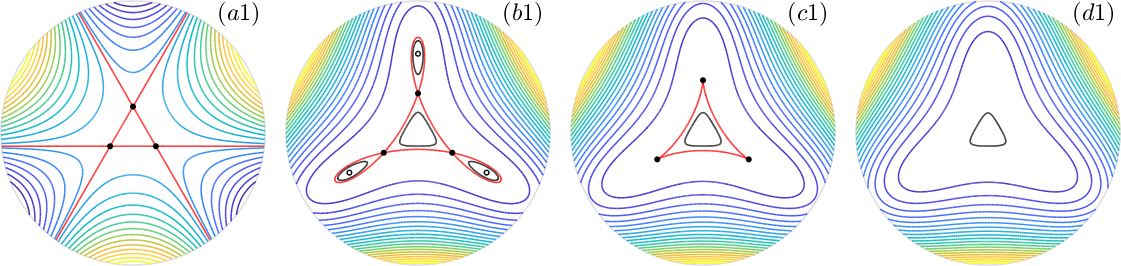} }}\quad
    \subfloat{{\includegraphics[width=0.85\textwidth]{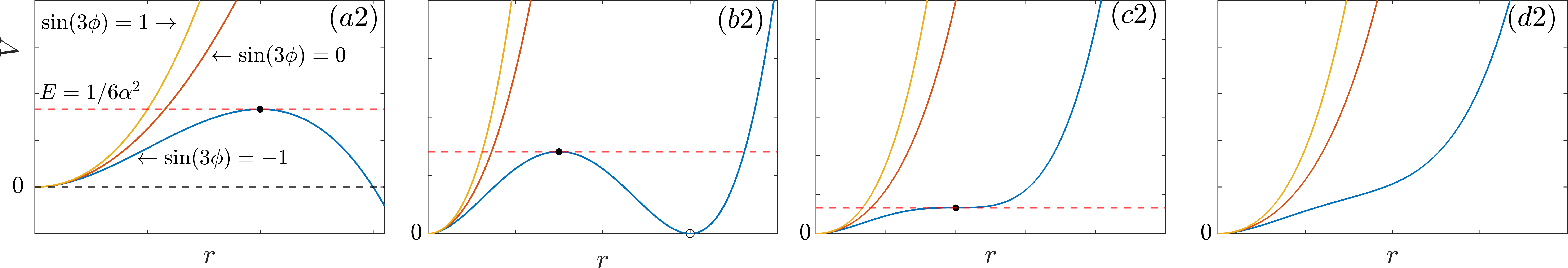} }}
    \caption{Equipotentials (upper panel) and schemes of potential landscape (lower panel), for $\alpha$-FPUT and the general case of different couplings $\lambda$, in polar coordinates. The stable and unstable (saddle) fixed points are denoted by open and full (black) circles. ($a1$-$a2$) $\alpha$-FPUT, this is exactly the well-known H\'enon-Heiles equipotential, where the line shown in red is the equipotential curve corresponding to the escape (saddle point) energy, $E_s=1/6\alpha^2$. ($b1$-$b2$) For $\lambda=1/18$, all equipotentials are confined, but below the critical  (saddle point) energy indicated by red (dashed) line, the equipotentials are composed of disconnected parts. ($c1$-$c2$) For $\lambda=1/16$, all equipotentials are confined and connected, but there exists a critical energy $E_s=1/3$ where saddles still exist. ($d1$-$d2$) For $\lambda=3/40$, all saddles disappear.}%
    \label{fig:epot1}%
  \end{figure*}
  \subsection{The classical Hamiltonian and the structure of the potential well}
  \label{sec3.1}

  The three-particle FPUT system with PBC is governed by Hamiltonian from Eq. \eqref{eq:fpu-pbc} with $N=3$.  From the transformation defined in Eq. \eqref{eq:inverse-odd}, one can easily prove that Hamiltonian of $\alpha$-FPUT takes the form
\begin{align}
  \mathcal{H}_\alpha =\frac{1}{2}\sum_{k=0}^2 P_k^2 +\frac{3}{2}\sum_{k=1}^2Q_k^2+\frac{3\alpha}{\sqrt{2}}(Q_1Q_2^2-\frac{1}{3}Q_1^3), 
\end{align}
and the quartic term of $\beta$-FPUT 
\begin{align}
  \sum_{j=1}^3\frac{\beta}{4}(x_{j+1}-x_j)^4= \frac{9\beta}{8}(Q_1^2+Q_2^2)^2. 
\end{align}
Note that $Q_0$ locating the center of mass is absent from $\mathcal{H}_\alpha$, implying the center of mass moves with constant momentum $P_0$. Then transforming to the center of mass frame and setting $t=\tau/\sqrt{3}, Q_1=\sqrt{2} q_2, Q_2=\sqrt{2}q_1$, we obtain \cite{ford1992fermi} canonically equivalent Hamiltonians
\begin{align}
    \label{eq:helon-heiles}
    H_\alpha = \frac{1}{2}\sum_{i=1}^2(p_i^2+q_i^2)+\alpha(q_1^2q_2-\frac{1}{3}q_2^3)
\end{align}
for $\alpha$-FPUT, which is exactly the H\'enon-Heiles Hamiltonian \cite{henon1964applicability} and for $\beta$-FPUT
\begin{align}
    \label{eq:beta-FPUT}
    H_\beta=\frac{1}{2}\sum_{i=1}^2(p_i^2+q_i^2)+\frac{3\beta}{4}(q_1^2+q_2^2)^2.
\end{align}
For the general three-particle FPUT, we set the scale as $t=\tau/\sqrt{3}, Q_1=\sqrt{2}q_2/\alpha , Q_2=\sqrt{2} q_1/\alpha$ and get
\begin{align}
    \label{eq:alpha-beta-FPUT}
    H = \frac{1}{2}\sum_{i=1}^2(p_i^2+q_i^2)+ q_1^2q_2-\frac{1}{3}q_2^3+ \lambda(q_1^2+q_2^2)^2,
\end{align}
where we denote the  coupling parameter of the quartic term 
\begin{align}
  \lambda=3\beta/(4\alpha^2).
\end{align} 
In polar coordinates $q_1=r\cos\phi$, $q_2=r\sin\phi$, the Hamiltonian of three-particle $\alpha$-FPUT is
  \begin{align}
      H_\alpha = \frac{1}{2}(p_r^2+\frac{\dot p_\phi^2}{r^2})+\frac{1}{2}r^2+\frac{\alpha}{3}r^3\sin3\phi,
  \end{align}
  with momenta $p_r=\dot r$, $p_\phi=r^2\dot\phi$. The associated equation of motion in polar coordinates are
  \begin{align}
      \dot p_r = -r-\alpha r^2\sin3\phi, \quad \dot p_\phi=\alpha r^3\cos3\phi.
  \end{align}
For the general FPUT given in Eq. \eqref{eq:alpha-beta-FPUT}
  \begin{align}
      \dot p_r = -r - r^2\sin3\phi-4\lambda r^3, \quad \dot p_\phi = r^3\cos3\phi.
  \end{align}

  The fixed points of the Hamiltonian flow are the points $(r,p_r,\phi,p_\phi)_m$ where all four derivatives for canonical motions are equal to zero. Thus, there are three unstable fixed (saddle) points depending on $\alpha$ for $\alpha$-FPUT
  \begin{align}
    \label{eq:phi_m}
      (r,p_r,\phi,p_\phi)_m=(1/\alpha,0,\phi_m,0),
  \end{align}
  with $\phi_m=\pi/2, 7\pi/6,11\pi/6$, and the energy at saddles $E_s=1/6\alpha^2$, also known as the escape energy. For the general case, the fixed points are
  \begin{align}
    \label{eq:lambda_pm}
      (r,p_r,\phi,p_\phi)_m=(\lambda_\pm,0,\phi_m,0),\ \lambda_{\pm}=\frac{1\pm\sqrt{1-16\lambda}}{8\lambda},
  \end{align}
  where $(\lambda_-,0,\phi_m,0)$ are saddles for couplings $\lambda \le 1/16$, $(\lambda_+,0,\phi_m,0)$ are stable fixed points if $\lambda< 1/16$. There is no unstable fixed point for $\lambda> 1/16$, while the point of origin is stable fixed point in all cases.

In Fig. \ref{fig:epot1} we show the equipotentials and schemes of potential landscape, for both $\alpha$-FPUT and the general case,  in which all the fixed points (except the point of origin) are identified. These fixed points are the classical origins of the so-called excited state quantum phase transitions \cite{cejnar2021excited} (ESQPT), which will be revealed in the study of density of states in Sec. \ref{sec5}. Clearly, the structure of equipotentials has $C_{3v}$ symmetry with respect to the rotation of $\phi$, due to the algebraic form of the classical Hamiltonians. Thus we see the $2\pi/3$ rotation symmetry and the reflection symmetry.

  \subsection{The Poincar\'e maps and SALI plots}
  \label{sec3.2}
\begin{figure}[h]
  \centering
  \includegraphics[width=1.0\linewidth]{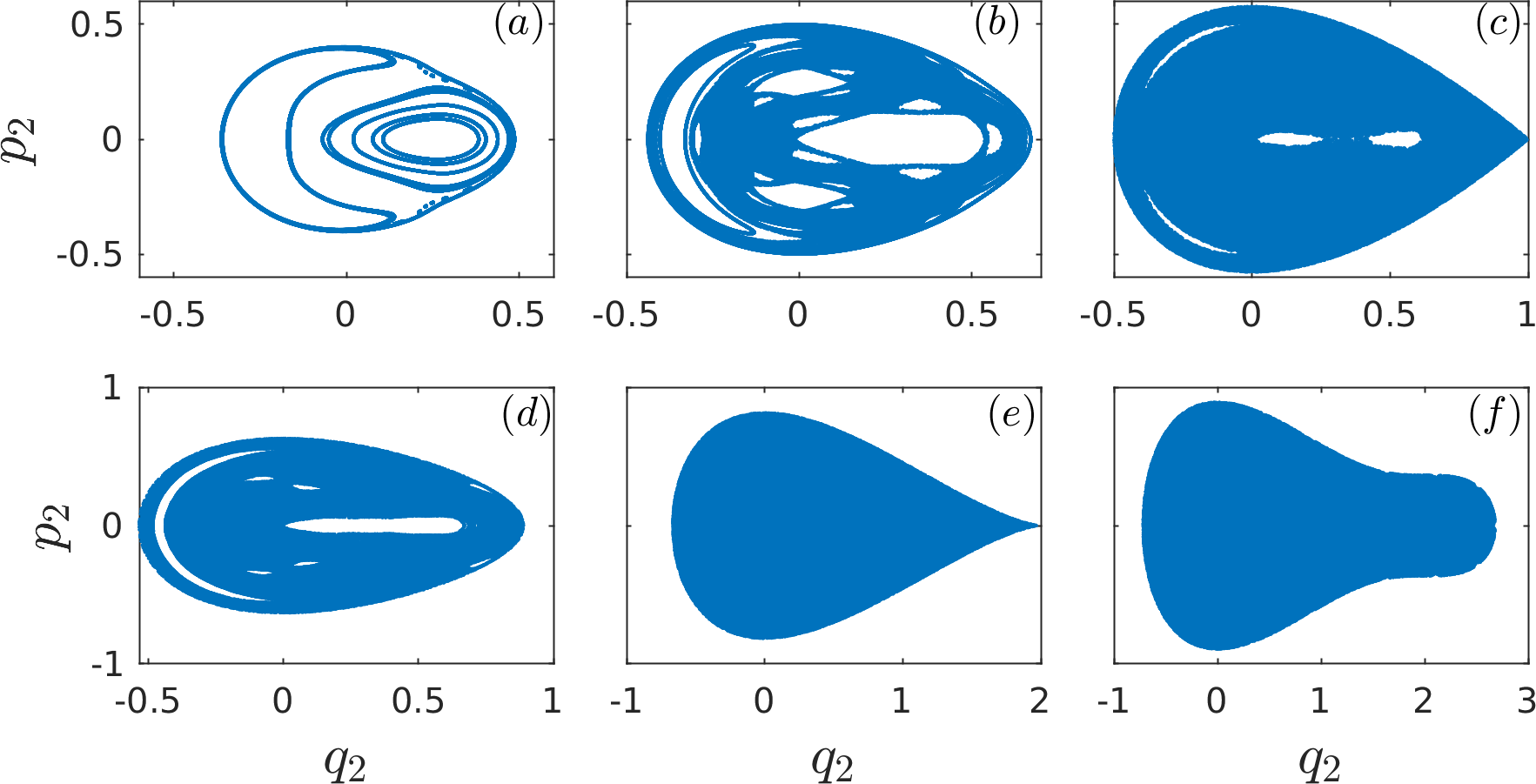}
  \caption{Classical SOS $(q_2,p_2)$ on $q_1 =0$ plane generated from 10 (random) orbits for different energies. The top panel is for three-particle $\alpha$-FPUT (with $\alpha=1$): ($a$) For energy $E=0.5E_s$ at the threshold energy, the closed curves correspond to regular classical orbits. ($b$) For energy $E=0.75E_s$ above the threshold energy. ($c$) For energy $E=E_s$ at the escape energy. ($d$-$f$) in the bottom panel are sections for $\lambda=1/16$ of the general FPUT,  at three energies $E=0.2, 1/3, 0.4$, while $E=E_s=1/3$ is the energy of  the saddles.}
  \label{fig:fput-pss} 
\end{figure}

  We integrate the equations of motion using the Adams-Moulton solver \cite{hairer1993multistep}, and present in Fig. \ref{fig:fput-pss} various classical SOS $(q_2,p_2)$ on $q_1 =0$ plane  for three energies, of both $\alpha$-FPUT with $\alpha=1$ and the general FPUT with $\lambda=1/16$. It was found that for $\alpha$-FPUT, i.e. the well-known H\'enon-Heiles model, below the order-chaos threshold energy $E=0.5E_s$  obtained from the concavity-convexity analysis of Toda \cite{toda1974instability}, all orbits lie on well-defined two-dimensional invariant tori in the four-dimensional phase space, as shown in Fig. \ref{fig:fput-pss}($a$). Above the threshold energy, there are chaotic orbits which fill a three-dimensional volume of phase space, 
as shown in  Fig. \ref{fig:fput-pss}($b$)($c$).

\begin{figure*}
  \centering
  \includegraphics[width=0.75\linewidth]{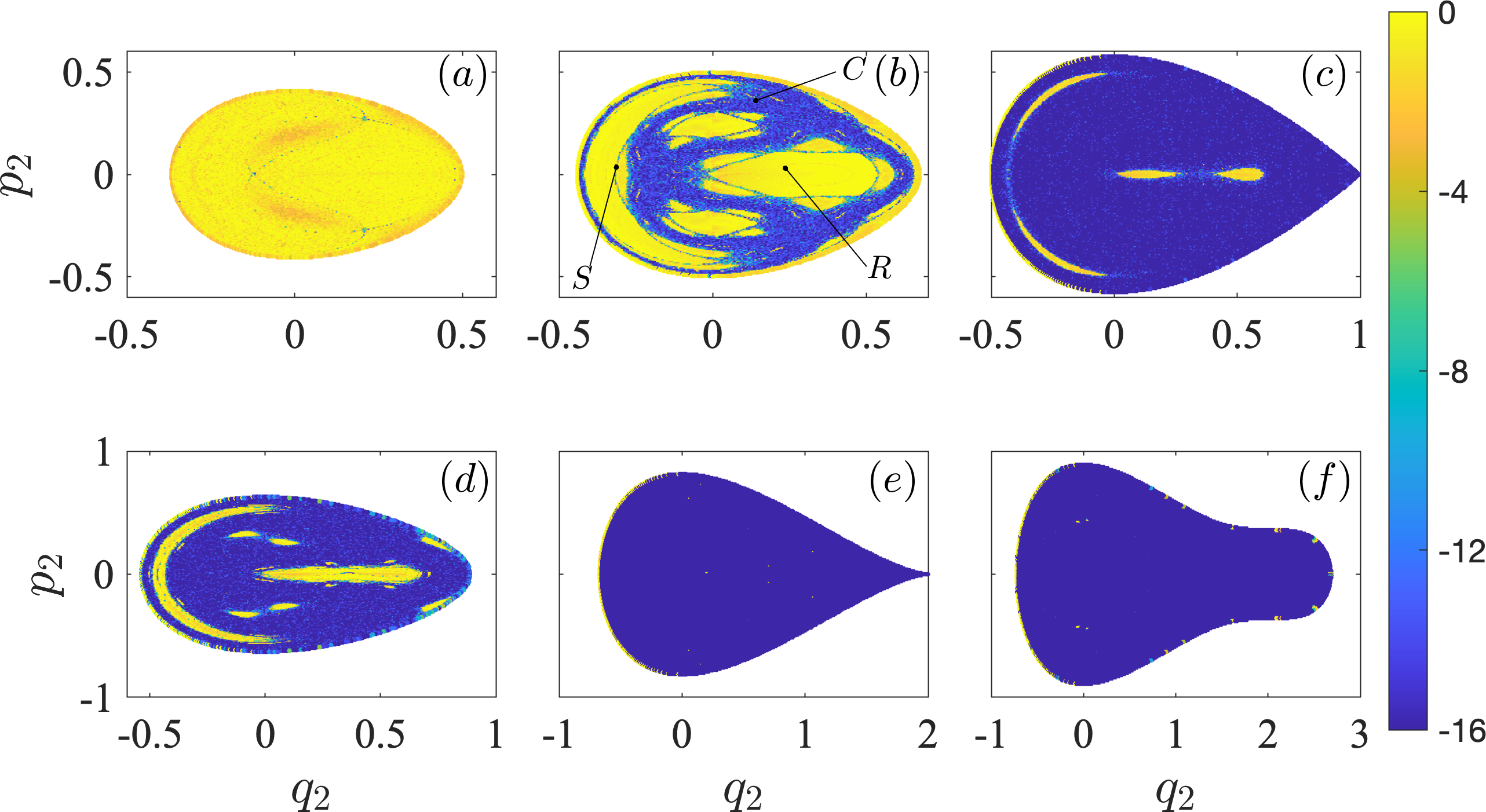}
  \caption{Regions of different (logarithmic) values of the SALI on the SOSs the same as ($a$-$f$) in Fig. \ref{fig:fput-pss}, at $t=1000$. The initial conditions colored dark blue correspond to chaotic orbits, the yellowish indicates ordered motion, and the intermediate suggests sticky orbits.}
  \label{fig:sali-pss}
\end{figure*}

Let $\chi_c(\textbf{q},\textbf{p})$ denote the characteristic function of the chaotic component, which takes the value of 1 on chaotic region and zero otherwise. To numerically compute $\chi_c$ across the energy surface, we employ the smaller alignment index (SALI) method \cite{bountis2012complex,skokos2016chaos}. This method relies on the evaluation of deviation vectors from a given orbit,  based on the equations of the tangent map obtained by linearizing the difference equations of a symplectic map as the following
\begin{align}
    \dot{\textbf{w}}=\Omega \cdot \nabla^2H(\textbf{x})\cdot \textbf{w}, \quad \Omega = \big[\begin{smallmatrix}
        0 & -I_d  \\
        I_d & 0  \\
        \end{smallmatrix}\big],
\end{align}
where $\textbf{x}=(\textbf{q}, \textbf{p})$ and the deviation vector $\textbf{w}=\delta \textbf{x}$, $\nabla^2 H(\textbf{x})$ is the Hessian matrix, $I_d$ being $d$-dimensional identity matrix.

In Fig. \ref{fig:sali-pss} we present a dense grid of initial conditions $(q_2,p_2)$ from the same SOSs as ($a$-$f$) in Fig. \ref{fig:fput-pss}, where the value of SALI of each point is plotted using an assigned color accordingly. Upon comparing with the corresponding SOS in Fig. \ref{fig:fput-pss}, we can have a much more detailed picture of the regions where chaotic or regular motion occurs, and at the borders between these regions we find intermediate colors which correspond to sticky orbits. 
\begin{figure}
  \centering
  \includegraphics[width=0.9\linewidth]{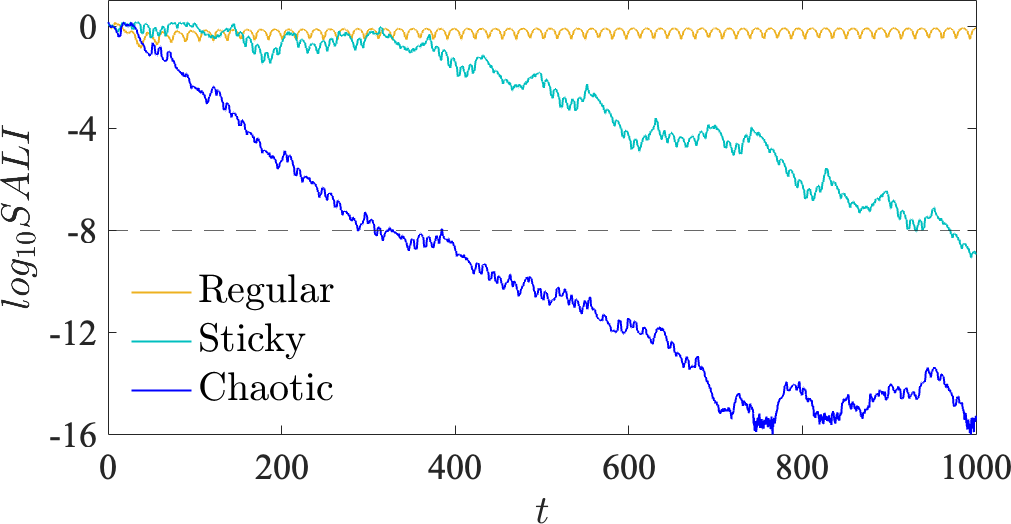}
  \caption{Evolution of the SALI with respect to time $t$, for three initial conditions chosen from panel ($b$) in Fig. \ref{fig:sali-pss} denoted by R (regular), S (sticky) and C (chaotic), while the integration time step $\Delta t$=0.05. The SALI of sticky orbit takes longer time about $t=900$ to approach the threshold value (it sticks to the border between the chaotic and regular region up to about $t=400$) than the chaotic case, which quickly reaches the threshold and below at about $t=300$.}
  \label{fig:sali-log}
\end{figure}
Fig. \ref{fig:sali-log} illustrates a typical time-evolution of the SALI,  for three types of orbits: the regular, the sticky ones, and the chaotic, it shows that the sticky orbit takes a considerable longer integration time than the chaotic ones to reveal its true chaotic nature. Our criterion for the classification of initial conditions belonging to a chaotic region is that SALI $\le 10^{-8}$ at $t=1000$, implying that the deviation vectors have been aligned. Here, the unit of (dimensionless) time is one period of the linear oscillator.

  \subsection{The relative Liouville volume measure of the chaotic part of the phase-space}
  \label{sec3.3}
 Hamiltonian systems with classically mixed-type dynamics, as exemplified in the Poincar\'e maps and SALI plots in Sec. \ref{sec3.2}, have regular quasi-periodic motion on $d$-dimensional invariant tori for some initial conditions ($d$ the degrees of freedom), and chaotic motion for the complementary initial conditions. The energy shell is therefore filled by regular and chaotic orbits. We define the chaotic fraction $\mu_c$  as the relative Liouville volume measure of the chaotic part of the phase-space as \cite{meyer1986theory,batistic2010semiempirical}
  \begin{align}
      \mu_c = \frac{\Phi_c}{\Phi}=\frac{\int d\textbf{q}d\textbf{p}\chi_c(\textbf{q},\textbf{p})\delta(E-H(\textbf{q},\textbf{p}))}{\int d\textbf{q}d\textbf{p}\delta(E-H(\textbf{q},\textbf{p}))},
  \end{align}
  where $\Phi$ is the phase space volume of the entire energy surface, and $\Phi_c$ the Liouville phase space volume of the chaotic region. The chaotic fraction $\mu_c$ can be used as an indicator of chaos, measures the transition from integrable dynamics with $\mu_c=0$ to the fully chaotic (ergodic) dynamics $\mu_c=1$. 
  
  From a statistical viewpoint, the chaotic fraction $\mu_c$ can be approximated by the ratio of the number of chaotic orbits $N_c$ signified by the SALI to the number $N_a$ of all orbits, in a large enough sample chosen from the entire energy surface, 
  \begin{align}
      \mu_c = \frac{\Phi_c}{\Phi} \approx \frac{N_c}{N_a}.
  \end{align}
   The numerical precision of $\mu_c$ would be improved by increasing $N_a$ (randomly selected from the whole available phase space). It should be noted that $\mu_c$ is not the relative area of the chaotic region in SOS, but of the volume in the full phase space. For the relation between these two quantities, see Refs. \cite{meyer1986theory,batistic2010semiempirical}.
  \begin{figure}
      \includegraphics[width=1.0\linewidth]{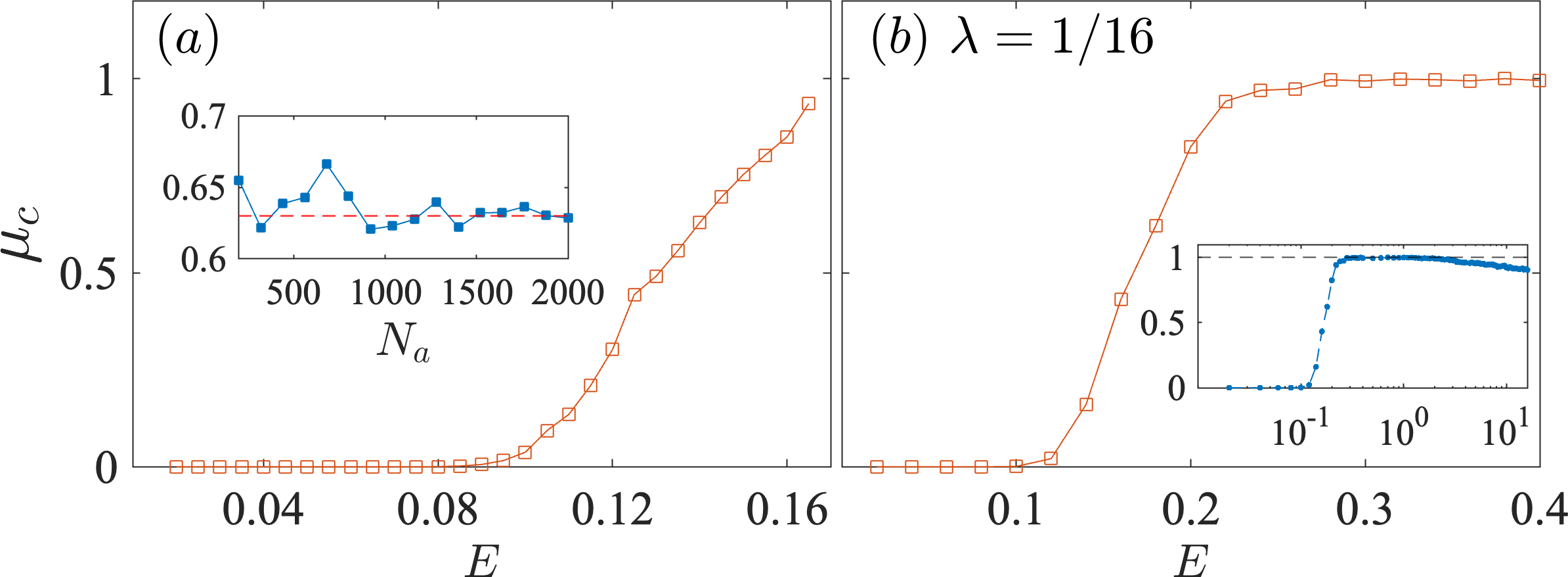}
      \caption{Chaotic fraction $\mu_c$ of the classical phase space as a function of energy $E$ for ($a$) the $\alpha$-FPUT with $\alpha=1$ (the inset shows at $E$=0.14, the convergence of $\mu_c$ with respect to $N_a$ the total number of samples), and ($b$) the general case with $\lambda=1/16$, the inset demonstrates that the value of $\mu_c$ will finally decrease at much higher energies, where the quartic term of potential starts to dominate.}
      \label{fig:fraction}
  \end{figure}

  The dependence of the chaotic fraction $\mu_c$ on energy for $\alpha$-FPUT and the general case of $\lambda=1/16$ is shown in Fig. \ref{fig:fraction}. In both cases we set $N_a=2000$ for which the numerical computation  of $\mu_c$ has converged, as illustrated in the inset of Fig. \ref{fig:fraction} ($a$). It shows that $\mu_c$ is zero below the threshold energy $E=0.5E_s$ for $\alpha$-FPUT, and increases monotonically with respect to the energy above the threshold, until the escape energy. The dependence of $\mu_c$ on the energy for the general case, shown in Fig. \ref{fig:fraction}($b$), is more complicated: it stays zero below a threshold energy, then increases monotonically until it saturates to $\mu_c=1$ for a while, meaning that the chaotic orbits  fill the entire available phase space, and finally it slowly decreases at higher energies, where the quartic term of the potential starts to dominate, exhibiting a transition from chaos to order.

  \section{The quantization of the three-particle FPUT system}

  \label{sec4}

  \subsection{Quantization and introduction of rotated bosonic operators}
  \label{sec4.1}
  The fact that $C_{3v}$ symmetry exists in the classical Hamiltonian, is more evident in the polar coordinates than in the Cartesian coordinates as we have discussed in Sec. \ref{sec3.1}. To take a better advantage of this symmetry, we define $\hat{q}_\pm = \hat{q}_1\pm i\hat{q}_2, \
  \hat{p}_{\pm}=\hat{p}_1\pm i\hat{p}_2$, and introduce the rotated bosonic operators \cite{karimi2014radial}
  \begin{equation}
  \label{eq:rotated-bosonic}
  \begin{aligned}
     a_\pm &= \frac{1}{\sqrt{2}}(a_1\mp i a_2)=\frac{1}{2\sqrt{\hbar}}(\hat{q}_\mp +i\hat{p}_\mp) ,\\
     a_\pm^\dagger &= \frac{1}{\sqrt{2}}(a_1^\dagger \pm i a_2^\dagger)= \frac{1}{2\sqrt{\hbar}}(\hat{q}_\pm - i\hat{p}_\pm), 
  \end{aligned}
  \end{equation}
  where the annihilation and creation operator fulfill the canonical commutation relations $[a_i,a_j^\dagger]=\delta_{ij}$, with $i,j\in\{1,2,\pm\}$. We can then check that for number operator $\hat{n}$ and angular momentum operator $\hat{\ell}=(\hat{q}_1\hat{p}_2-\hat{q}_2\hat{p}_1)/\hbar$,
  \begin{equation}
  \begin{aligned}
      &\hat{n}=a_1^\dagger a_1+a_2^\dagger a_2=a_+^\dagger a_++ a_-^\dagger a_-,\\
      &\hat{\ell}= i(a_2^\dagger a_1 -a_1^\dagger a_2)=a_+^\dagger a_+-a_-^\dagger a_-.
  \end{aligned} 
  \end{equation}
  
The quantized Hamiltonian of $\alpha$-FPUT 
  \begin{align}
      \hat{H}_\alpha=\hbar(\hat{n}+1)-i\alpha(\hat{q}_+^3-\hat{q}_-^3)/6,
  \end{align}
  and for the general case
  \begin{align}
      \hat{H}=\hbar(\hat{n}+1)-i(\hat{q}_+^3-\hat{q}_-^3)/6+\lambda(\hat{q}_+\hat{q}_-)^2.
  \end{align}
  \subsection{Matrix elements in the circular two-mode basis}
  \label{sec4.2}
  As $[\hat{n},\hat{\ell}]=0$, $\hat{n}$ and $\hat{\ell}$ possess simultaneous eigenfunctions $ |n,l\rangle$, which we refer to as the circular two-mode basis, with $l=-n,-n+2,\cdots,n\  (n\in \mathbb{N}_0)$,  $n$ the radial quantum number and $l$ the orbital angular momentum.  One can then express the annihilation and creation operator in this circular basis, as well as the operators $\hat{q}_\pm$ (see Appendix \ref{appA} for details). Consequently, the matrix elements of the cubic coupling terms are given as
  \begin{align}
      \label{eq:coeff-cubic}
    \langle n',l'|\hat{q}_\pm^3|n,l\rangle=\hbar^{3/2}\delta_{l',l\pm3}\sum_{m\in\mathcal{M}_1} k_m^\pm (n,l)\delta_{n',n+m},
  \end{align}
  where $\mathcal{M}_1=\{\pm 1, \pm3\}$, $k_m^+(n,l)=k_m^-(n,-l)$ with
  \begin{equation}
  \begin{aligned}
      \label{eq:no-phase}
      &k_{-1}^+=3\sqrt{(n-l)(n-l-2)(n+l+2)/8}, \\
      &k_{1}^+=3\sqrt{(n-l)(n+l+2)(n+l+4)/8},  \\
      &k_{-3}^+=\sqrt{(n-l)(n-l-2)(n-l-4)/8},  \\
      &k_{3}^+=\sqrt{(n+l+2)(n+l+4)(n+l+6)/8}.
  \end{aligned}
\end{equation}
For the quartic coupling term, we have
  \begin{align}
    \langle n',l'|(\hat{q}_+\hat{q}_-)^2|n,l\rangle=\hbar^2\delta_{l'l}\sum_{m\in\mathcal{M}_2}k_m(n,l)\delta_{n',n+m}
  \end{align}
  with $\mathcal{M}_2=\{0, \pm 2, \pm4\}$, and the coefficients
  \begin{equation}
  \begin{aligned}
      &k_0=\frac{3}{2}n^2-\frac{1}{2}l^2+3n+2, \\
      &k_{-2}= n\sqrt{n^2-l^2},\ k_2=(n+2)\sqrt{(n+2)^2-l^2},\\
      &k_{-4}=\frac{1}{4}\sqrt{n^2-l^2}\sqrt{(n-2)^2-l^2},\\
      &k_4=\frac{1}{4}\sqrt{(n+2)^2-l^2}\sqrt{(n+4)^2-l^2}.
  \end{aligned}
\end{equation}

So in the circular basis $|n,l\rangle$, the cubic terms in quantum $\alpha$-FPUT couple states with $\Delta l=\pm 3$. Consequently, the coupling takes place in three decoupled sets of basis states: $\{a\}$ $l\in \{\dots -6, -3 ,0, 3, 6\dots\}$, $\{b\}$ $l\in \{\dots -5, -2 ,1, 4\dots\}$ and $\{c\}$ $l\in \{\dots -4, -1 ,2, 5\dots\}$. Each set has non-degenerate eigenvalues, set $\{a\}$ the singlet maps onto itself under time reversal and  set $\{b\}$ maps onto  set $\{c\}$ and \emph{vice versa}. Due to the time reversal symmetry of the FPUT Hamiltonian, set $\{b\}$ and  set $\{c\}$ are doublets and have the same eigenspectra \cite{brack1993quantum}.  The number of states $\mathcal{N}$ (the size of the Hilbert space composed of both the singlet and doublets) for fixed $N$ as the cutoff of $n$, is $(N+1)(N+2)/2$, the dimensions of subspace of the singlets $\mathcal{N}_S$ and $\mathcal{N}_D$ of the doublets are 
  \begin{align}
    \label{eq:subspace-dim1}
    \begin{cases}
        \mathcal{N}_S =\mathcal{N}_D =\frac{(N+1)(N+2)}{6},
        \ \text{if } N\not\equiv 0\pmod3,\\
        \mathcal{N}_S =\frac{N(N+3)}{6}+1,\ \mathcal{N}_D =\mathcal{N}_S-1, \text{otherwise.}
    \end{cases}
\end{align} 

The quartic term in the  three-particle quantum $\beta$-FPUT does not introduce any coupling between different $l$, due to the conservation of angular momentum.
 \subsection{The role of symmetries}
 \label{sec4.3}
 The property of eigenspectra can also be verified from the $C_{3v}$ symmetry of the classical Hamiltonian with respect to the rotation of $\phi$, that they must belong to the irreducible representations of the point symmetry group $C_{3v}$ (Mulliken notation): the subspaces of two doublets are of $E$ symmetry, and the singlet is a combination of ($A_1$, $A_2$) symmetry. In Appendix \ref{appB} we give more details about the state representation and symmetries, based on a new basis defined as $|n,l,s\rangle$ \cite{weissman1982quantum1} 
 \begin{align}
    |n,l,s\rangle=a_{l,s}(|n,l\rangle +s|n,-l\rangle)/\sqrt{2}, \quad s=\pm 1,
    \end{align}
where $a_{l,s}=s^{\text{mod}(l,3)}$ and $l\ge 0$. It should be noted that here the basis states $|n,0,+1\rangle$ are not normalized.
The Hilbert space can then be divided further into four decoupled subspaces: 
 \begin{itemize}
     \item subspace 1: spanned by the basis set \{$|n,l,1\rangle, \mod(l,3)=0$\},
 \item subspace 2: spanned by the basis set \{$|n,l,-1\rangle, \mod(l,3)=0$\},
   \item subspace 3: spanned by the basis set \{$|n,l,1\rangle, \mod(l,3)\ne 0$\},
 \item subspace 4: spanned by the basis set \{$|n,l,-1\rangle, \mod(l,3)\ne 0$\}.
 \end{itemize}
 
 From this classification of subspaces, it is clear that the basis set of subspace 1 and 2 combined is the singlet we have defined, while the set of subspace 3 and 4 is the set of two doublets. Subspace 1 is of $A_1$ symmetry, subspace 2 is of $A_2$ symmetry, and these two subspaces are distinct, where the states with $l=0$ do not exist in subspace 2. Subspace 3 and 4 are of $E$ symmetry, and have the same eigenspectra since the matrix elements are identical, which are the eigenspectra of the doublets.

  \section{The density of states}
  \label{sec5}
  The fixed points of classical Hamiltonian we discussed in Sec. \ref{sec3.1} correspond directly to the singularity which appears in the derivative of smoothed DOS and in energy densities of various observables, whose quantum analogs are also referred to as ESQPT, widely studied in various quantum systems such as Lipkin-Meshkov-Glick
  (LMG) model, the Dicke model \cite{brandes2013excited,bastarrachea2014comparative}, and the interacting boson systems. In the following, we give the analytical result of the smoothed DOS and its derivative for both $\alpha$-FPUT and the general case, using the Thomas-Fermi rule. Then by using the method
  of quantum typicality  \cite{hams2000fast,bartsch2009dynamical,elsayed2013regression,jin2021random}, we verify that the obtained quantum density of states, as well as the ones calculated by the Krylov subspace method, perfectly agree with the smoothed DOS.
  \begin{figure*}
    \centering
    \subfloat{{\includegraphics[width=0.33\textwidth]{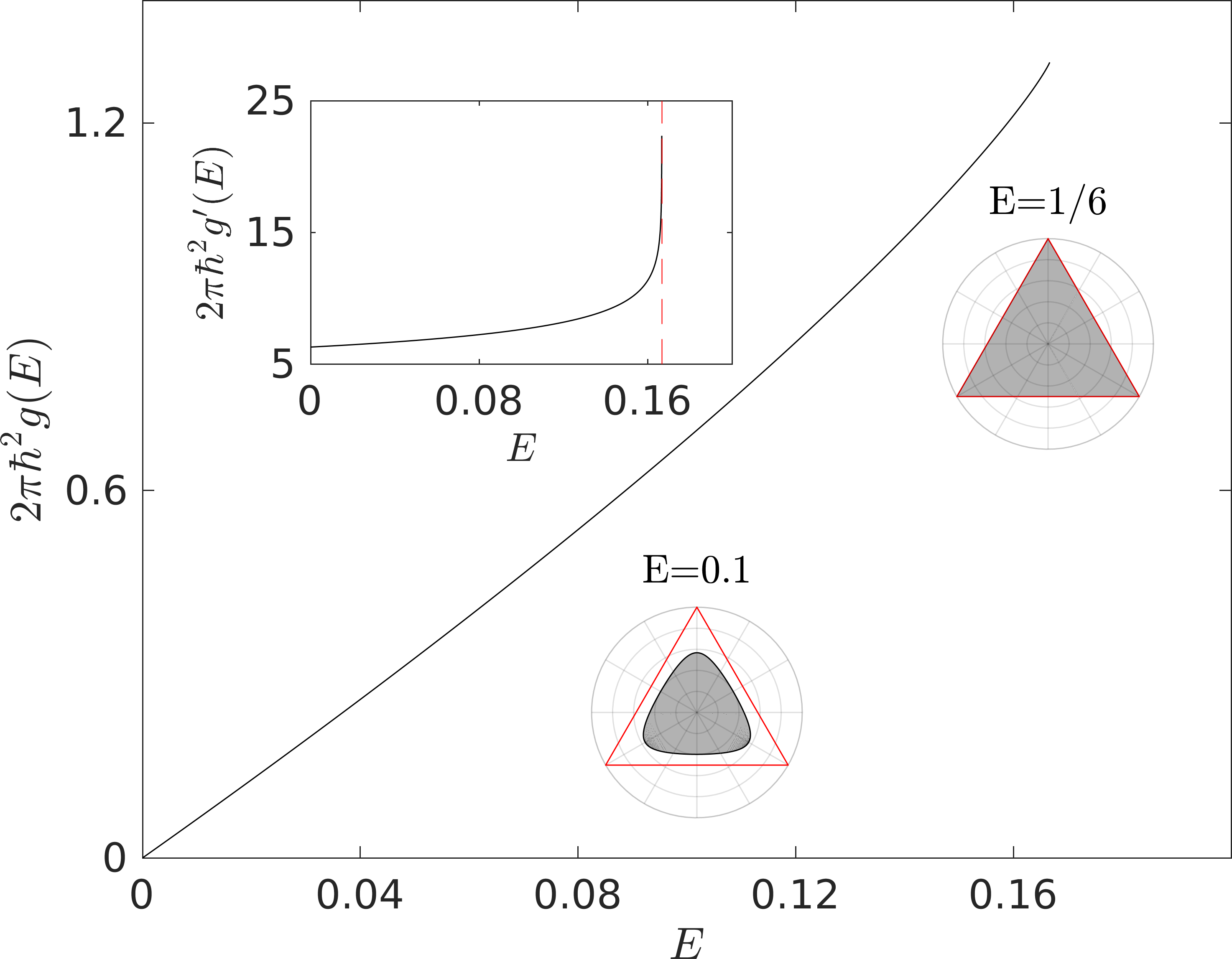} }}
    \subfloat{{\includegraphics[width=0.31\textwidth]{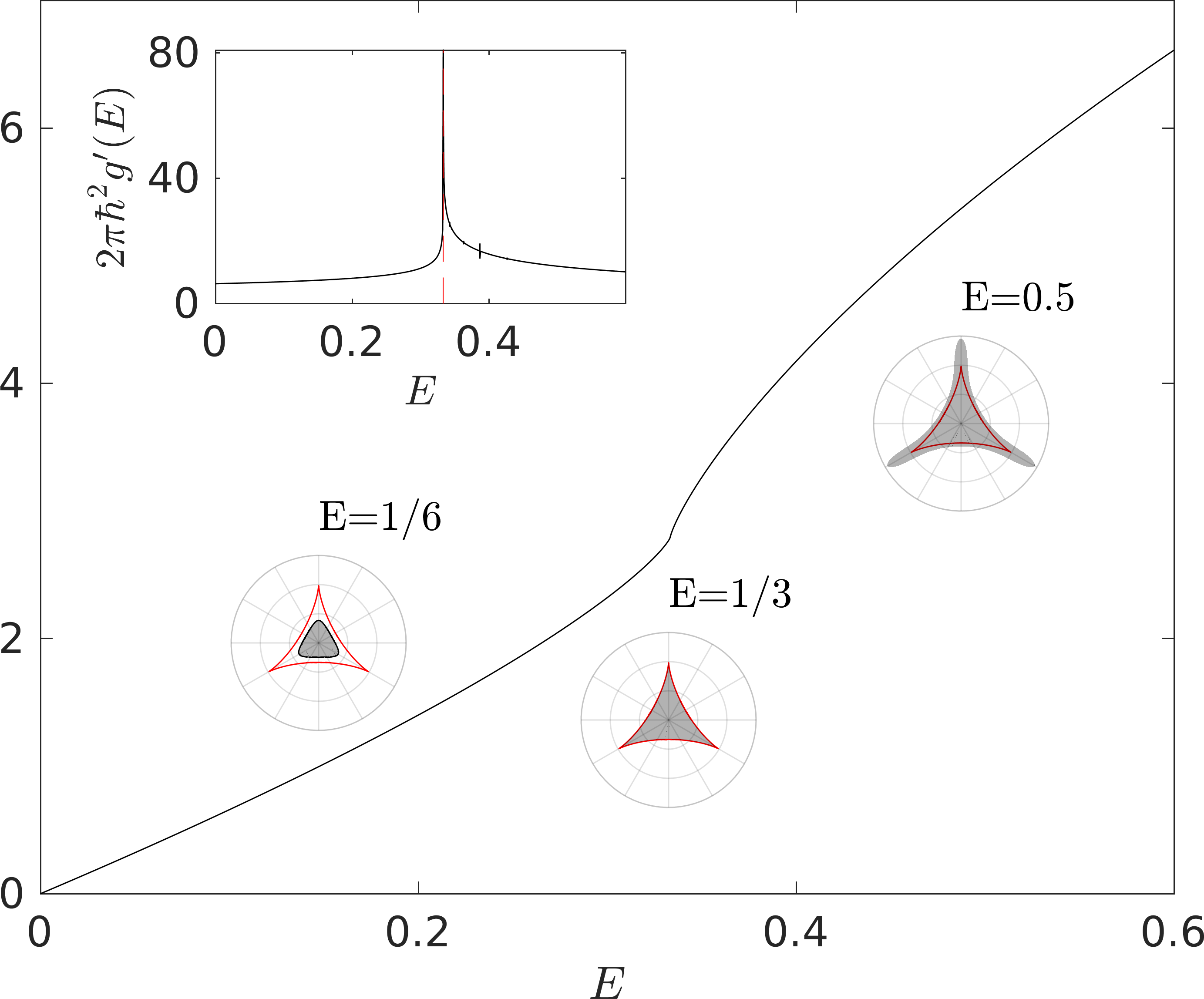} }}
    \subfloat{{\includegraphics[width=0.31\textwidth]{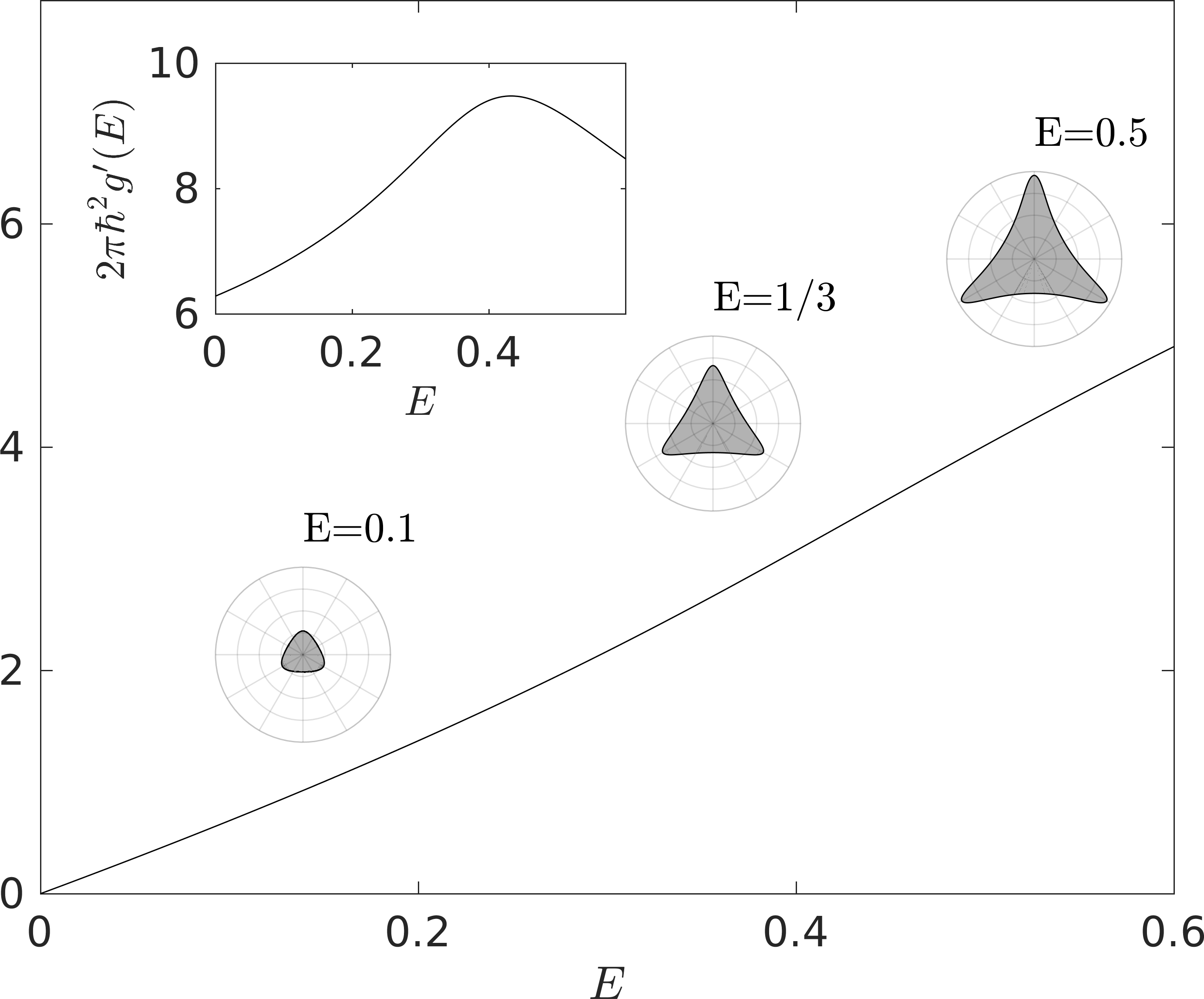} }}
    \caption{Scaled available phase space volume $2\pi \hbar^2 g(E)$ for the three-particle $\alpha$-FPUT ($\alpha=1$) as a function of energy $E$ (left), the general case with $\lambda=1/16$ (middle), and $\lambda=3/40$ (right). The derivatives of scaled $g(E)$ are shown as inset at the top left, where the red dash indicates the energy of saddles $E=E_s$. Polar plots of the available phase space (gray zones) for different energies are shown in the bottom and top right.}
    \label{fig:TF1}
  \end{figure*}
  \subsection{The Thomas-Fermi rule}
  \label{sec5.1}
  The Thomas-Fermi expression for DOS is given as 
  \begin{align}
      g(E)=\frac{1}{(2\pi\hbar)^2}\int d\textbf{q}d\textbf{p}\delta(E-H(\textbf{q},\textbf{p})),
  \end{align}
  which is the volume of the available classical phase space for a given energy $E$ divided by $(2\pi\hbar)^2$. $g(E)$ is also referred to  as the semiclassical approximation (smooth component) of the quantum density of states, obtained from the contribution of zero-length orbits in path integral, according to Gutzwiller's trace formula \cite{gutzwiller1971periodic,berry1976closed}. As a consequence of the quadratic momentum in the Hamiltonians, the integral can be simplified (see Appendix \ref{appC}),
  \begin{align}
    \label{eq:TF-vol}
    g(E)=\frac{1}{2\pi\hbar^2}\int dq_1dq_2=\frac{1}{2\pi\hbar^2}\int r drd\phi,
\end{align}
given the constraint $V(r,\phi)\le E$. Thus, a further simplification of the integral depends on solving $V(r,\phi)=E$ for $r(\phi)$ the root. For $\alpha$-FPUT of the H\'enon-Heiles potential, bounded solution exists when $E\le 1/6\alpha^2$, 
\begin{align}
  \begin{cases}
  r_+(\phi) = \frac{1}{\alpha\sin 3\phi}(\cos\frac{\theta}{3}-\frac{1}{2}), &\text{if } \sin3\phi \ge0, \\
  r_-(\phi)=\frac{-1}{\alpha\sin 3\phi}(\cos\frac{\theta+\pi}{3}+\frac{1}{2}), &\text{if }\sin3\phi \le0,
  \end{cases}
\end{align}
where $\cos\theta = 12E\alpha^2\sin^23\phi-1$, $\theta\in[0,\pi]$. The resulting Thomas-Fermi expression is then
\begin{align}
  g(E) =\frac{3}{4\pi\hbar^2}\big[\int_0^{\pi/3}r_+^2d\phi+\int^{2\pi/3}_{\pi/3}r_-^2d\phi\big].
\end{align}

From the evaluation of $g(E)$ we can calculate the derivative $g^\prime(E)$ and obtain
\begin{align}
  g^\prime(E)=\frac{\sqrt{2}}{\pi\hbar^2}\int^{\pi}_{\theta_E}\frac{\sin(\frac{\theta}{6}+\frac{\pi}{3})}{\sqrt{12E\alpha^2-1-\cos\theta}}d\theta,
\end{align}
where $\theta_E$ is given by $\cos\theta_E=12E\alpha^2-1$.  Writing the scaled energy $E=1/6\alpha^2-\delta$ with $\delta\to 0_+$, this integral for $g'(E)$ can be approximated by Legendre's incomplete elliptic integral of the first kind, from which we extract explicitly in Appendix \ref{appC} the logarithmic singularity 
\begin{align}
  \label{eq:henon-log}
  g^\prime(\frac{1}{6\alpha^2}-\delta) \approx -\frac{c}{\sqrt{2}\pi\hbar^2}\ln \frac{3\delta \alpha^2}{8},
\end{align}
where $c$ is a constant that fulfills $\sqrt{3}/2<c<1$. 

\begin{figure}
  \includegraphics[width=0.95\linewidth]{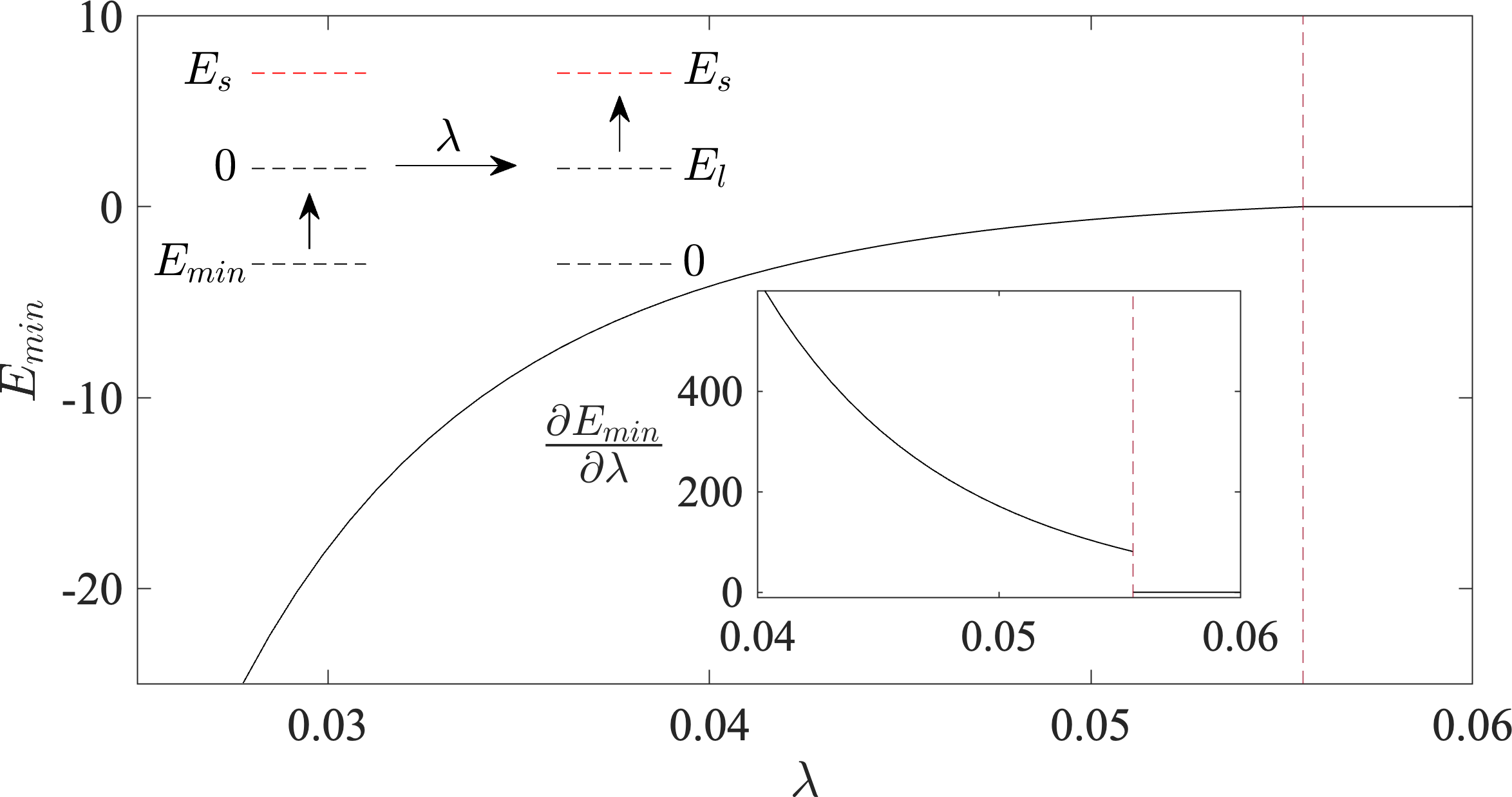}
  \caption{Energy minimum $E_{min}$ as a function of the quartic coupling $\lambda$ , the vertical (dashed) line indicates $\lambda=1/18$. The inset (bottom right) shows the first-order derivative of $E_{min}(\lambda)$, while the inset on the top left is a scheme for two $\lambda$-regions: $0<\lambda<1/18$ (left) and $1/18<\lambda<1/16$ (right).}
  \label{fig:minimum}
\end{figure}

\begin{figure*}
  \centering
  \subfloat{{\includegraphics[width=0.32\textwidth]{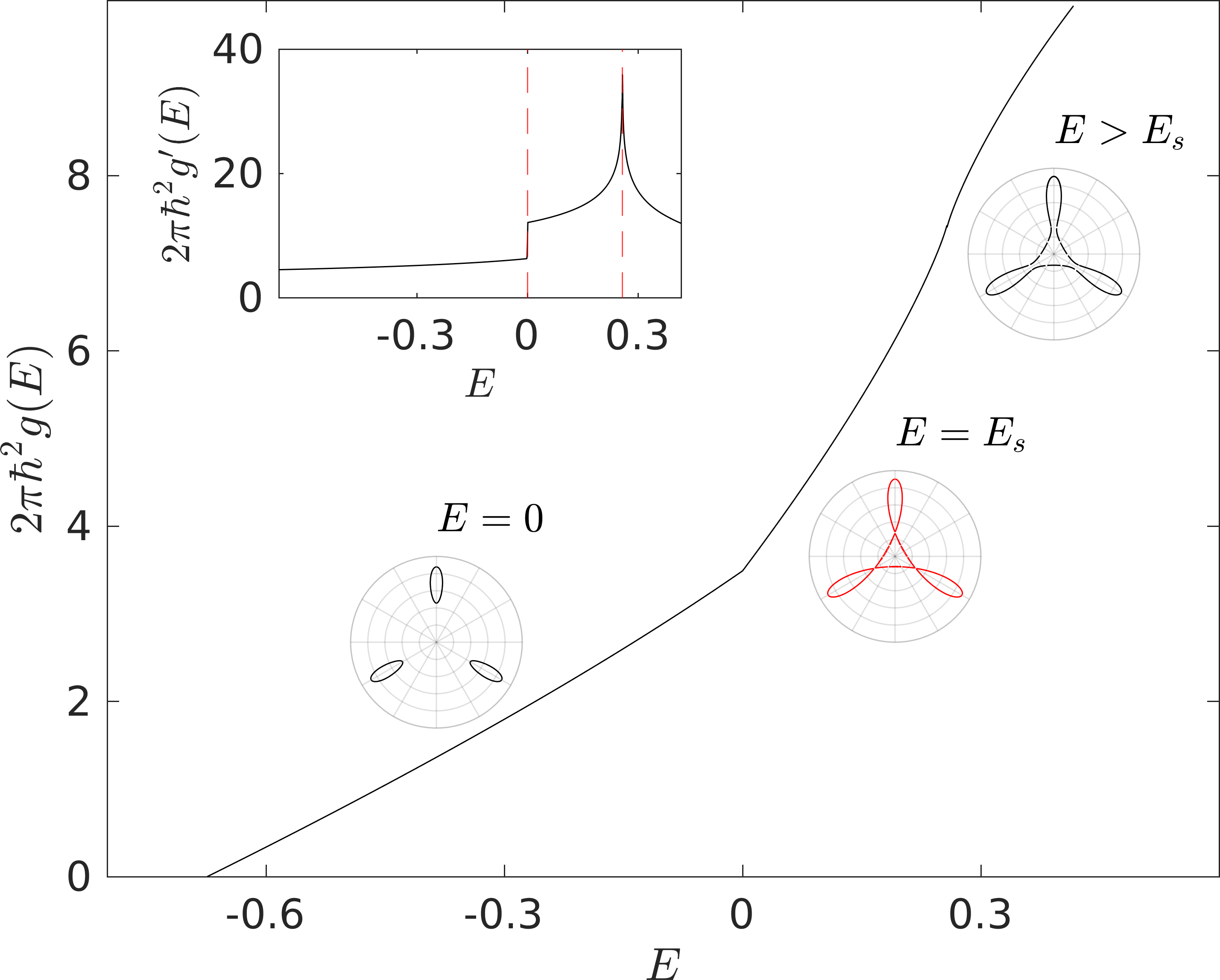} }}%
  \subfloat{{\includegraphics[width=0.31\textwidth]{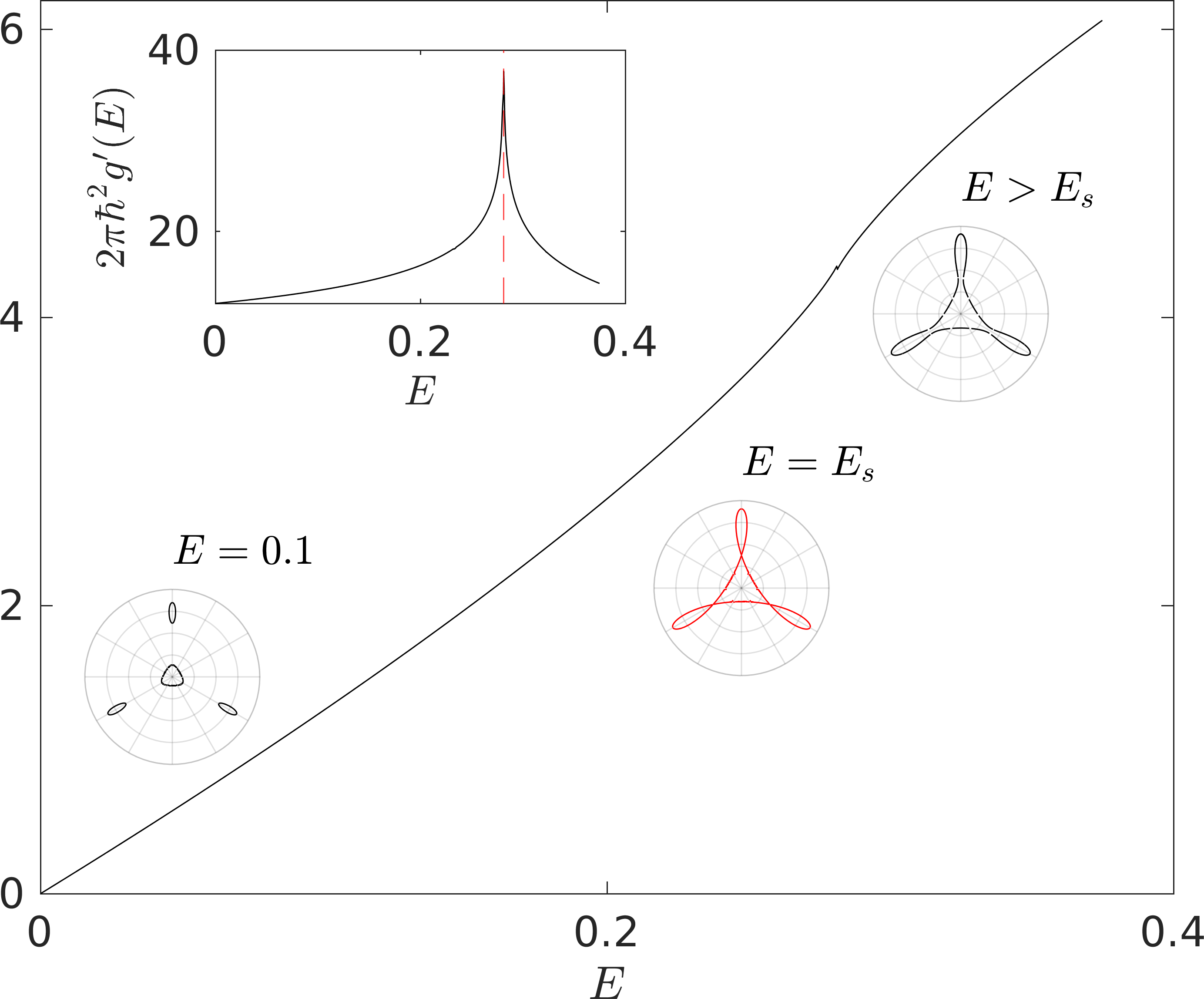} }}%
  \subfloat{{\includegraphics[width=0.31\textwidth]{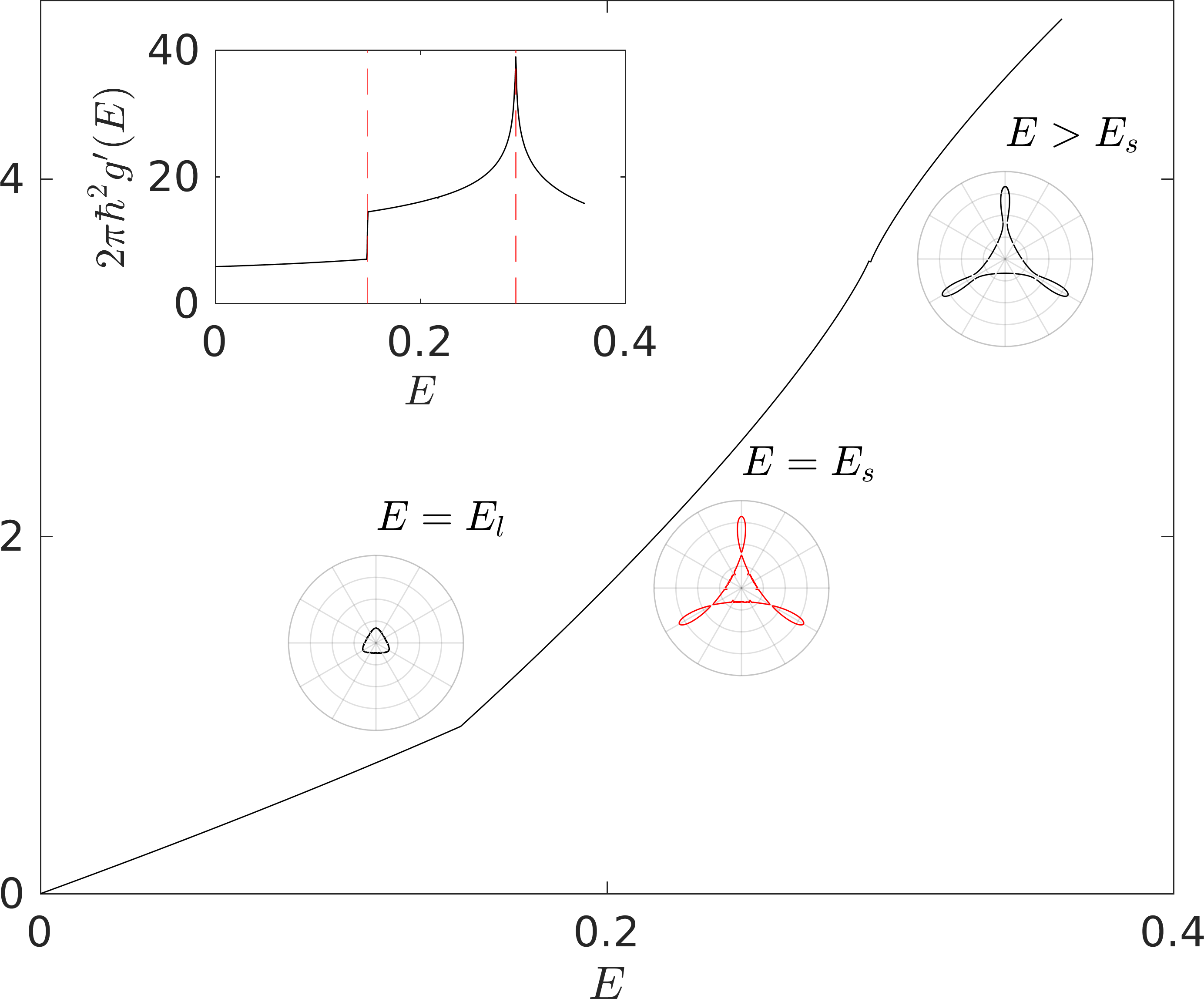} }}%
  \caption{Scaled available phase space volume $2\pi \hbar^2 g(E)$ for the general three-particle FPUT as a function of energy $E$, for $\lambda=1/20$ (left), $1/18$ (middle) and $3/52$ (right).  The derivative of scaled $g(E)$ is shown as inset at the top left, the red dashes indicate the critical energies at classical fixed points. The available phase space for different energies are shown as polar plots.} 
  \label{fig:TF2}
\end{figure*}
For the general case, the integral in Eq. \eqref{eq:TF-vol} depends on solving the quartic equation 
\begin{align}
  \label{eq:quartic1}
 E=V(r,\phi)=\frac{1}{2}r^2+\frac{1}{3}r^3\sin3\phi+\lambda r^4.
\end{align}
An analytical representation of the solution of the quartic potential could be found. Examining the potential landscapes illustrated in Fig. \ref{fig:epot1} and supported by discriminant analysis, it is evident that for $\lambda\ge1/16$, there exists only one real positive root, which can be represented by hyperbolic functions.  In Appendix \ref{appC} we have derived the analytical expression of this root,  from which one can directly evaluate the integral in Eq. \eqref{eq:TF-vol} and then the derivative.

In Fig. \ref{fig:TF1} we plot the scaled available phase space volume $2\pi\hbar^2g(E)$ and its derivative as a function of energy, also polar plots of the available phase space, for $\alpha$-FPUT with $\alpha=1$, the general case with $\lambda=1/16$ and $\lambda=3/40$. The derivative shows logarithmic divergence  at $E=E_s=1/6$ for $\alpha$-FPUT as we have extracted in Eq. \eqref{eq:henon-log}, and at $E=E_s=1/3$ for the general case with $\lambda=1/16$, where $E_s$ is the energy of saddles. This behavior can be understood by looking at the geometry (polar plots) of the available phase space: when $E$ approaches $E_s$, the saddles emerge and the boundary of the available phase space is not smooth anymore.  For $\lambda=3/40 > 1/16$, there are no evident discontinuities, because there are no saddles.

 For $\lambda<1/16$, multiple roots would emerge below the saddle energy,  accordingly, the parameter space of $\lambda$ can further be partitioned into two distinct $\lambda$-regions: $0<\lambda<1/18$ and $1/18<\lambda<1/16$. In the first region the three potential wells are the deepest, while in the second region the central well is the deepest, at $\lambda=1/18$ all four stable minima are the same (see Fig. \ref{fig:epot1}$(b2)$). Therefore, in the first region, as the energy increases, a notable transformation occurs in the geometry of the available phase space: it evolves from three disconnected parts to four disconnected parts, ultimately culminating in an interconnected configuration. Conversely, in the subsequent region ($1/18<\lambda<1/16$), there is a transition from a single connected part to four disconnected parts, before eventually returning to the interconnected. 

 The global minimum of energy for $\lambda<1/16$ is the minimum energy of all stable fixed points,  defined as $E_{min}=\min\{
  V(\lambda_+,\phi_m), 0\}$. The local minimum energy $E_l$  between the stable fixed points and the energy of the saddles $E_s$ are given as
  \begin{align}
 E_l=\max\{
     V(\lambda_+,\phi_m), 0\},\ E_s=V(\lambda_-,\phi_m),
   \end{align}
where  $E_{\min}\le E_l<E_s$, $\lambda_\pm$ and $\phi_m$ are defined in Eqs. \eqref{eq:phi_m}-\eqref{eq:lambda_pm} in Sec. \ref{sec3.1} . In Fig. \ref{fig:minimum} we show $E_{min}$ as a function of $\lambda$, and a scheme of two $\lambda$-regions in one inset,  where the other inset reveals that there is a discontinuity of the first derivative of $E_{min}$ at $\lambda=1/18$, a signature of so-called first-order quantum phase transition. The analytical solution of the quartic equation in Eq. \eqref{eq:quartic1} varies with respect to energy, in two $\lambda$-regions,  as we have clarified in Appendix \ref{appC}, where the explicit expressions of $g(E)$ are given.
 
 The scaled available phase space volume and the derivatives as a function of $E$  are plotted in Fig. \ref{fig:TF2} for three quartic couplings. For $\lambda=1/20<1/18$, the energy at the three stable fixed points (around the point of origin) are the global minimum, while $E=0$ the energy at the point of origin is a local minimum above $E_{min}$, showing that there is a step discontinuity  of $g'(E)$ at $E=E_l=0$, and a logarithmic divergence at $E=E_s$. For $\lambda=1/18$ there is no stable fixed point above $E_{min}=0$, therefore no step discontinuities occur. When $\lambda=3/52$ ($1/18<\lambda<1/16$), it shows that at $E_l>0$ there is a step discontinuity of $g'(E)$, and at $E=E_s$ the logarithmic divergence occurs. These changes can be well understood by looking at the geometry of the available phase space of each $\lambda$-interval (see the polar plots in Fig. \ref{fig:TF2}): when energy goes across $E_l$, the number of disconnected parts changes, while the boundary remains smooth, but as energy approaches $E_s$, the boundary of the available phase space is not smooth.

\subsection{The method of quantum typicality}
  \label{sec5.2}
  The method of quantum typicality is an approximation scheme based on typicality, in which approach the DOS is obtained by solving the time-dependent Schr\"odinger equations, through implementing the Chebyshev expansion techniques \cite{de2004computational,weisse2006kernel}, followed by the fast Fourier transformation (FFT) of the retarded Green's function. Given the density of states of a quantum system
  \begin{align}
      \rho(E)=\sum_n \delta(E-E_n)=\frac{1}{2\pi}\int_{-\infty}^{+\infty}dt\  e^{iEt}\text{Tr}e^{-i
  \hat{H}t},
  \end{align}
  where $\hat{H}$ is the Hamiltonian of the quantum system and  $n$ runs over all eigenvalues of $\hat{H}$ (in this subsection we put $\hbar=1$). The trace in the integral can be estimated accurately by sampling over random vectors
  \begin{align}
      \frac{1}{\mathcal{N}}\text{Tr}e^{-i\hat{H}t}\approx\langle\psi(0)|\psi(t)\rangle =\langle \psi(0)|e^{-i\hat{H}t}|\psi(0)\rangle,
  \end{align}
  where $\mathcal{N}=\text{dim}\{\hat{H}\}$ denotes the dimension of the Hilbert space and $|\psi(0)\rangle$ is a random state drawn according to the Haar measure, the error scales with $1/\sqrt{\mathcal{N}}$. Here $|\psi(t)\rangle$ can effectively be calculated by making use of the Chebyshev polynomial algorithm for the matrix exponential. 
  
  The quantum DOS then can be approximated by 
  \begin{align}
      \label{eq:dos-typ}
      \rho(E)\approx \frac{\mathcal{N}}{2\pi}\int^{+T}_{-T}e^{iEt}\langle \psi(0)|\psi(t)\rangle dt,
  \end{align}
where $T$ denotes the time range required to obtain the energy resolution $\pi/T$. To cover the full range of eigenvalues, a sampling interval $\Delta t=\pi/||\hat{H}||$ (the 2-norm of $\hat{H}$) is sufficient according to the Nyquist sampling theorem, implying a restriction of the time steps that can be used. In practice, the integration over time in Eq. \eqref{eq:dos-typ} can be performed by the discrete Fourier transform as
  \begin{align}
      \rho(k\pi/T)\approx \frac{\mathcal{N}T}{2\pi M}\sum_{j=-M}^{M-1}e^{i\pi jk/M}\langle\psi(0)|\psi(j\tau)\rangle, 
  \end{align}
  with $k=-M,\cdots, M-1$, and $\tau=T/M <\Delta t$ is the time step at which we sample the inner product $\langle \psi(0)|\psi(t)\rangle$, where in evaluation one can use FFT.

  \subsection{Quantum density of states - numerical results}
  \label{sec5.3}
  In Sec. \ref{sec4.2} we give the matrix representation of the Hamiltonian for $\alpha$-FPUT and the general case in circular-mode basis. In calculations of quantum DOS we set $N=1200$. It turns out that the convergence of numerical solutions is better in circular-mode basis than in Cartesian-mode basis.  To obtain quantum density of states, generally exact diagonalization (ED) is required, which is only feasible for small Hilbert space. In this case $\mathcal{N}\sim 10^6$ is too large for ED, instead we employ the Krylov subspace method \cite{saad1981krylov,saad2003iterative} to obtain the energy spectra across different energy intervals, and also the method of quantum typicality to avoid the calculation of eigenvalues.

For quantum three-particle $\alpha$-FPUT, i.e. the quantum H\'enon-Heiles model, we set the Planck constant  $\hbar=3\times 10^{-4}$. The number of energy levels below the classical escape energy $E_s$ is less than one third of the whole truncated spectra, and numerically the energy eigenvalues closest to $E_s$ change less than one percent of the averaged level spacing as the cutoff $N$ is increased from 1200 to 1600, which means that the eigenvalues can accurately represent the spectra by the criterion of convergence. For the general case with $\lambda=1/16$, we set $\hbar=3\times 10^{-3}$, and make calculations up to less than 1/20 of the truncated spectra. In both cases, by Krylov subspace method, the averaged number of states are taken over $\Delta n=60$ consecutive levels in specific energy interval of width $\Delta E$. The quantum DOS is then in the mean defined as $\rho(E)=\Delta n/\Delta E$. In the typicality approach, we make the average over 10 initial random state. 

\begin{figure}
  \centering
  \subfloat{{\includegraphics[width=0.9\linewidth]{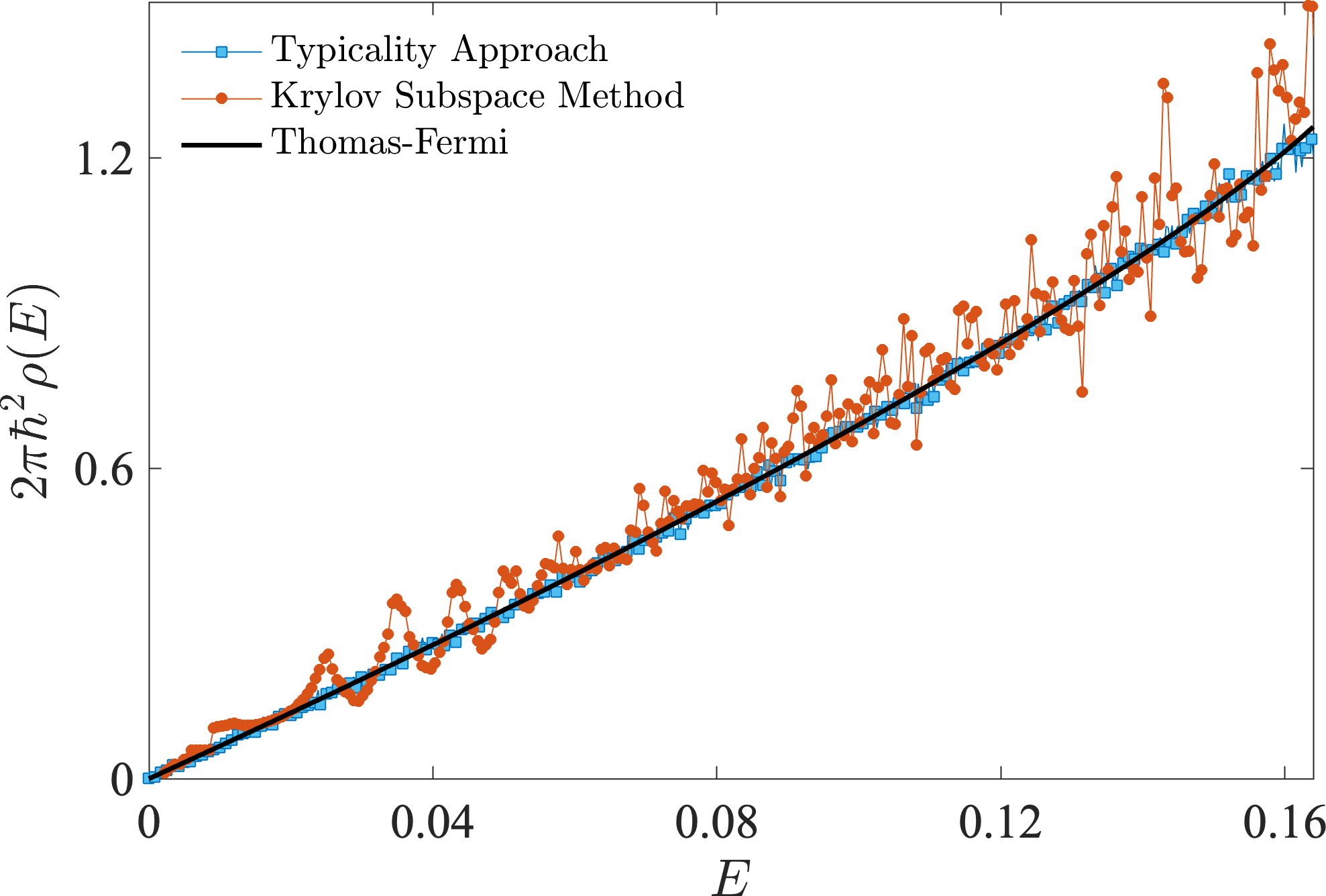} }}\\
  \subfloat{{\includegraphics[width=0.9\linewidth]{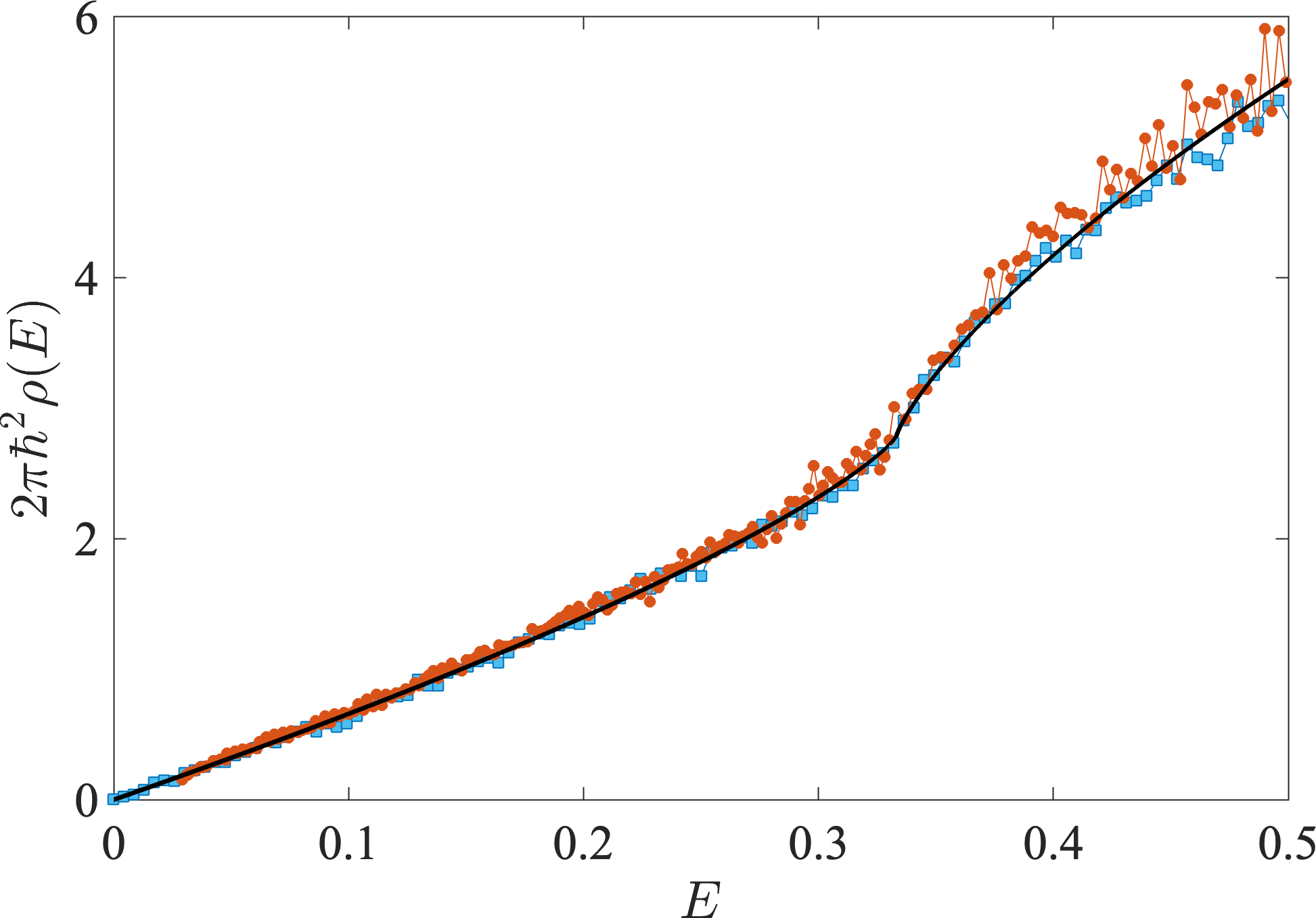} }}
  \caption{Rescaled quantum density of states $2\pi\hbar^2\rho(E)$ as obtained from the averaged number of state (solid circles), with $\rho(E)=\frac{\Delta n}{\Delta E}$ where $\Delta n=60$ is the number of states over energy interval $\Delta E$, and from the typicality approach (solid squares) averaged over 10 initial random states. The cutoff of n here is $N=1200$, providing $\mathcal{N}=721801$ for the calculation of the energy spectrum: we set $\hbar=3\times10^{-4}$ for $\alpha$-FPUT with $\alpha=1$ (top), and $\hbar=3\times 10^{-3}$ for the general case with $\lambda=1/16$ (bottom). They both agree in the mean with the Thomas-Fermi estimation (solid line).}
  \label{fig:DoS}
\end{figure}

The rescaled quantum density of states from two different methods are shown in Fig. \ref{fig:DoS} for both the $\alpha$-FPUT and the general case. They  agree with the Thomas-Fermi result and even capture the kink at saddle energy $E=E_s=1/3$ for the general case, especially the ones from the typicality approach. It is worth noting that the quantum DOS of $\alpha$-FPUT calculated in the mean exhibits prominent regular oscillations for energies $E<0.06$, referred to as quantum beats. They can be interpreted in terms of classical period orbits by a calculation of their amplitudes in the Gutzwiller trace formula (see Ref. \cite{brack1993quantum}). 

Our numerical results demonstrate that the semiclassical density of states correctly describes the quantum spectra of three-particle FPUT, and thus consequently can be used to perform the unfolding of the quantum spectra, which is essential for the spectral statistics. Conversely, it also can be used as a verification that the numerical quantum spectra are truly converged with respect to the cutoff $N$ of $n$. 

\section{The statistical properties of energy spectra}
\label{sec6}
\subsection{The transition to quantum chaos}
\label{sec6.1}
In Sec. \ref{sec3.3}, by calculating the chaotic fraction $\mu_c$ over the energy shell, we show in Fig. \ref{fig:fraction} a transition to classical chaos indicated by the variation of $\mu_c$ across the energy range. This transition would also be manifested in the quantum realm, arising from the quantum-classical correspondence. To study the transition to quantum chaos, first we use the nearest-neighbor-level spacing distribution $P(s)$ for short-range statistical properties of spectrum, which is the Poisson distribution $P(s)=e^{-s}$ for the quantum integrable case, and the Wigner-Dyson distribution $P(s)=(\pi/2)s\exp(-\pi s^2/4)$ from the random-matrix statistics for the quantum chaotic case with the time-reversal symmetry,  where $s$ denotes the spacing between two unfolded consecutive energy levels.  

To acquire spectral statistics across a wide range of energy, it is essential to have statistically reasonable number of energy levels  $N(E,\delta E)$ in a narrow energy interval around chosen energy $E$, as $[E-\delta E/2, E+\delta E/2]$ where $\delta E=\eta E$. The width of the energy interval needs to be narrow enough, $\eta \ll 1$, so that each energy level is associated with the same classical dynamics. For this purpose, we introduce a variation of the Planck constant depending on the energy as $\hbar(E)=(E/E_s)\hbar_c$, 
from a basic estimation
\begin{align}
  N(E,\delta E) \simeq g(E)\delta E = \frac{\eta\mathcal{A}}{2\pi}(E/\hbar)^2,
\end{align}
where from our analysis of smoothed DOS in Sec. \ref{sec5.1}, $\mathcal{A}=2\pi\hbar^2g(E)/E$ can be approximated to be a constant in the first order for $E\lesssim E_s$. For quantum $\alpha$-FPUT and the general case with $\lambda=1/16$, we set the cutoff of $n$ as $N=1800$. In the former case $\hbar_c=3\times 10^{-4}$ provides ten times more energy levels above its escape energy $E_s$, while in the latter case $\hbar_c=4\times 10^{-4}$ there are about two times more levels above the saddle energy $E_s$. The eigenstates below (and around) $E_s$ are thus well converged. In both cases, $\eta \simeq 0.01$ and the number of consecutive energy levels is $N(E,\delta E)= 1200$, calculated by the Krylov subspace method. 

\begin{figure}
  \includegraphics[width=1.0\linewidth]{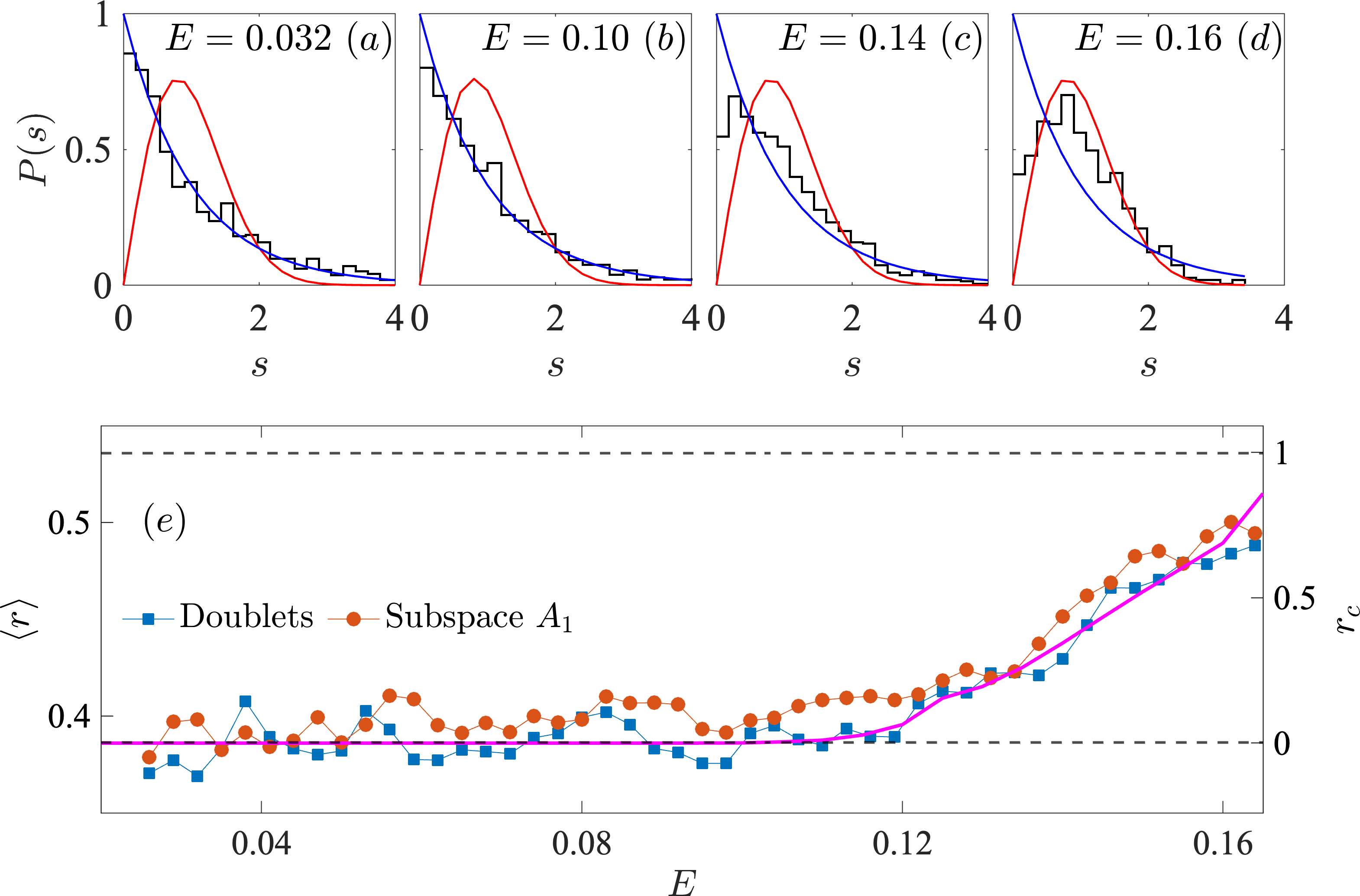}
  \caption{($a$-$d$) The histogram of neighbor-level spacing distribution $P(s)$ of doublets in quantum $\alpha$-FPUT ($\alpha=1$), for several energies with $N=1800$, the irreducible Hilbert space of size is $\mathcal{N}_D= 540900$. The Planck constant is set as $\hbar(E)=(E/E_s)\hbar_c$ where $\hbar_c=3\times 10^{-4}$ and $E_s$=1/6, to effectively calculate 1200 levels within the energy interval whose width is approximately $1\%$ of each energy. The red and blue solid curves in each panel show the Wigner-Dyson and Poisson distributions. ($e$) Mean spacing ratio $\langle r\rangle$ (left axis) and $r_c$ (right axis), where the solid line denotes the power function of classical fraction $f(\mu_c)=\mu_c^\kappa$ with $\kappa=2.3$, as a function of $E$,  with two horizontal (dashed) lines denoting $\langle r\rangle_W=4-2\sqrt{3}\approx 0.5359$ (upper) and $\langle r\rangle_P=2\ln2-1\approx 0.3863$ (lower).}
  \label{fig:ratio-henon}
\end{figure}

\begin{figure}
  \centering
  \includegraphics[width=1\linewidth]{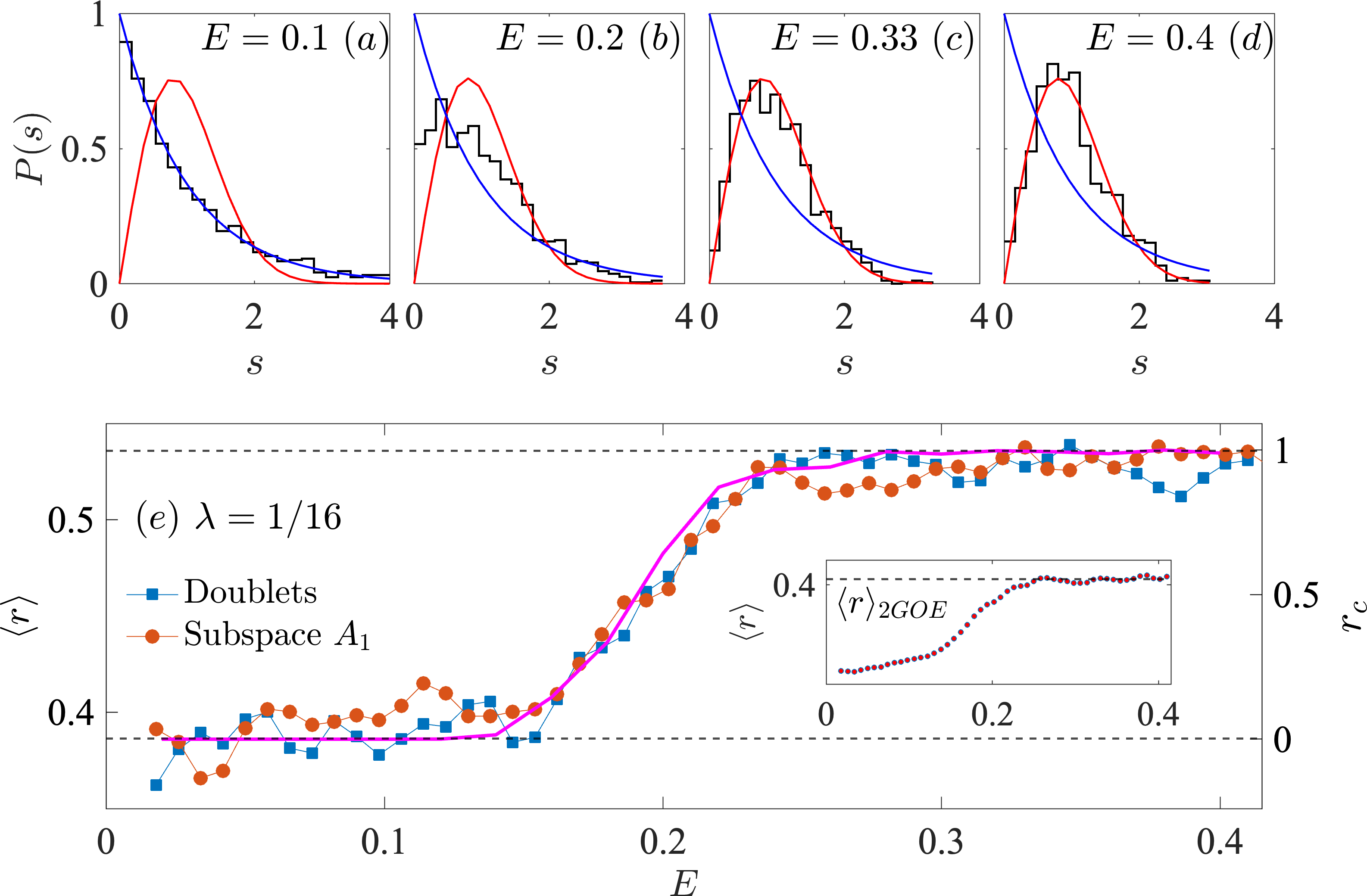}
  \caption{Analogous data as in Fig. \ref{fig:ratio-henon},  but now for the general case with $\lambda=1/16$, and the variational Planck constant is set as $\hbar(E)=(E/E_s)\hbar_c$, where $E_s=1/3$ is saddle energy and $\hbar_c=4\times 10^{-4}$. In panel ($e$) the inset is the mean spacing ratio of the singlets as a function of $E$, where the dashed line denotes $\langle r\rangle_{2GOE}\approx 0.4234$. }
  \label{fig:ratio-gen}
\end{figure}

We show in panels ($a$)-($d$) of Figs. \ref{fig:ratio-henon}-\ref{fig:ratio-gen} the behavior of $P(s)$ of the doublets for different energies. For the unfolding of partial energy spectra we have used the (ninth-order) polynomial fitting procedure. It exhibits a transition from Poisson distribution to Wigner-Dyson distribution as the energy increases. It should be noted that there is no time reversal symmetry in the doublets, but due to the mirror symmetry from the $C_{3v}$ symmetry, giving rise to another antiunitary symmetry, the doubly degenerate subspace is expected to show GOE statistics for energies where $\mu_c =1$ \cite{leyvraz1996anomalous,schafer2002transition,haake2018quantum}. Clearly, in the $\alpha$-FPUT case, it does not go exactly to the Wigner-Dyson distribution when energy increases to $E_s$, because unlike the general case, $\mu_c\lesssim 0.94$ for $E\le E_s$ indicates a mixed-type system below the saddle (escape) energy, for which we will treat the spectral statistics in more details  in Sec. \ref{sec6.2}.

To further illustrate the transition from regularity to chaos, we consider the spacing ratios, defined as \cite{atas2013distribution}
\begin{align}
  r_n = \min(\frac{s_n}{s_{n-1}},\frac{s_{n-1}}{s_n}),
\end{align}
with $s_n=E_n-E_{n-1}$ the consecutive level spacing, from $\{E_n\}$ an ordered set of energy levels. This quantity is now considered an effective tool to distinguish regular from chaotic quantum spectra, especially for quantum many-body systems, mostly due to its avoiding of numerical spectral unfolding, since ratios are independent of the local density of states. The approximate formulae for the distribution of $P(r)$ can be derived from random matrix statistics, the resulting mean spacing ratio $\langle r \rangle$ for the Poisson level spacing is $\langle r \rangle_p=2\ln2-1$, while for GOE statistics $\langle r \rangle_W=4-2\sqrt{3}$.  We can then define a quantity $r_c$ the normalized mean spacing ratio as an indicator of quantum chaos, namely
\begin{align}
r_c = \frac{\langle r \rangle-\langle r \rangle_P}{\langle r \rangle_W-\langle r \rangle_P},
\end{align}
obviously,  for the integrable $r_c=0$ while $r_c=1$ is for the fully chaotic.

In panel $(e)$ of Figs. \ref{fig:ratio-henon}-\ref{fig:ratio-gen}, we plot (on the left) $\langle r \rangle$ as a function of $E$, for both the doublets and subspace $A_1$ from the singlets. Of either subspace, $\langle r \rangle$  shows a transition from nearly Poisson value $\langle r \rangle_P$ to a value near $\langle r \rangle_W$ with increasing energy, up to $E_s$ for $\alpha$-FPUT, while for the general case it shows a transition to the GOE value $\langle r \rangle_W$ at energies where the classical chaotic fraction $\mu_c=1$. The route to quantum chaos is in general following the same transition to classical chaos as shown in Fig. \ref{fig:fraction}, but there exists a slight discrepancy with regard to the energy of the transition point of integrability breaking, where the energy in the quantum scenario exhibits a lag. One explanation is that this regime is not deep enough in the semiclassical limit and consequently the number of energy levels in the energy interval is just 1200.  The inset of Fig. \ref{fig:ratio-gen}($e$) shows the mean spacing ratio of the singlets, as a function of energy. For chaotic spectra with parity, as it has been pointed out \cite{giraud2022probing}, the value for $\langle r \rangle$ obtained from numerical simulations of random matrices is $\langle r\rangle_{2GOE}\approx0.4234$, which is clearly illustrated in our plot.

Additionally, on the right in panel $(e)$ of Figs. \ref{fig:ratio-henon}-\ref{fig:ratio-gen}, we depict a power function $f(\mu_c) = \mu_c^\kappa$ as a function of energy, with the power exponent $\kappa$ set to 2.3. From both plots, we see approximately there is $r_c=f(\mu_c)$. It should be noted that this is not an exact relation between the classical chaotic fraction $\mu_c$ and the indicator of quantum chaos $r_c$, but it reveals that there may exist a simple connection between the classical  and quantum indicator, for which we will derive the analytical expression in the upcoming paper \cite{yanrobnik3}.

\subsection{The Berry-Robnik-Brody distribution}
\label{sec6.2}

\begin{figure*}
  \centering
  \includegraphics[width=0.8\linewidth]{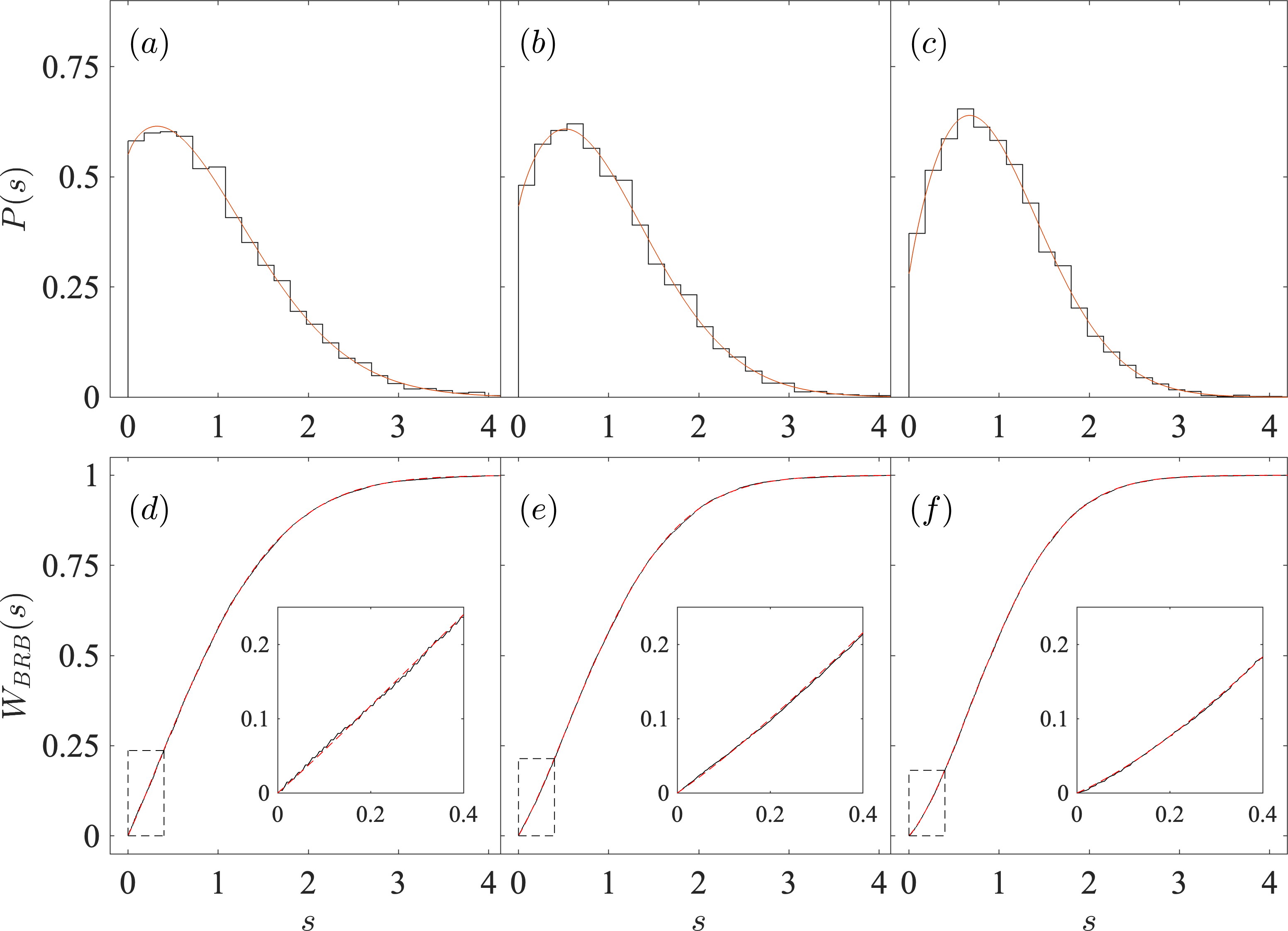}
  \caption{The histograms of level spacing distribution $P(s)$ (up panels) and cumulative level spacing distribution $W_{BRB}(s)$ (down panels) for three energy intervals $[E-\delta E/2, E+\delta E/2]$ with $\delta E/E\simeq 0.02$ at $E=0.14\ (a)(d), 0.15\ (b)(e), 0.16\ (c)(f)$, comprising about 7800 levels from the doublets of quantum $\alpha$-FPUT with $\alpha=1$. The best fitting BRB curves are (red) solid lines, and the inset magnifications display deviation of the numerical data from the best fitting BRB distribution. The fitting parameters $(\beta, \rho_c)$ from left to right are: (0.820, 0.673), (0.840, 0.757), (0.882, 0.853), while the classical fraction $\mu_c$ obtained from SALI is 0.630, 0.753 and 0.850, respectively. In both panels, we set the cutoff of $n$ as $N=2000$ and $\hbar=1.5\times 10^{-4}$, the resulting $\mathcal{N}_D\sim 1.3\times 10^6$.}
  \label{fig:HenonBRB}
\end{figure*}

From the transition to quantum chaos, we see that for intermediate energies in both the $\alpha$-FPUT and the general case, the level spacing distribution is between the Poisson and GOE statistics, since the classical fraction $\mu_c<1$ in the intermediate case, means classically a mixed-type system with structured phase space which has both regular and chaotic regions. Spectral statistics of this type was first theoretically treated by Berry and Robnik \cite{berry1984semiclassical}, where the classical chaotic fraction $\mu_c$ plays a crucial role, assuming that the chaotic states are uniformly extended over the classical invariant chaotic component. Further studies have shown that uniformity condition is not satisfied in a generic system if the dynamical (or quantum) phase space localization phenomenon occurs if $t_H < t_T$, where the Heisenberg time is $t_H= 2\pi\hbar \rho(E)$ with $\rho(E)$ being the quantum DOS, $t_T$ is the transport time controlling the rate of classical diffusion \cite{batistic2013dynamical,batistic2013quantum,robnik2023brief}. 

To extend the Berry-Robnik distribution such that the localization effects are included, Prosen and Robnik \cite{prosen1993energy,prosen1994semiclassical} have introduced the well-known (empirical) Brody distribution to describe the statistics of the energy spectra of the localized chaotic eigenstates
\begin{align}
  P_B(s)=c s^\beta \exp(-ds^{\beta+1}),
\end{align}
where from the normalization conditions $\int_0^\infty P_B(s)ds =1$ and  $\int_0^\infty sP_B(s)ds =1$ we obtain
\begin{align}
  c= (\beta+1)d, \quad d=\gamma^{\beta+1},
\end{align}
with $\gamma=\Gamma(\frac{\beta+2}{\beta+1})$,  $\Gamma(x)$ being the gamma function and $\beta\in[0,1]$: $\beta=0$ indicates the maximal localization in chaotic eigenstates and Poisson statistics, while $\beta=1$ corresponds to the uniformity and the GOE statistics. The degree of localization can be characterized by $\beta$, directly related to the ratio $t_H/t_T$ between two typical time scales.  For $t_H/t_T \gg 1$, the value of $\beta$ is expected to be close to 1, while  if $t_H/t_T \ll 1$, $\beta\to 0$. The corresponding gap probability is 

\begin{align}
  \mathcal{E}_B(s)=\frac{1}{\gamma(\beta+1)}Q\left(\frac{1}{\beta+1},(\gamma s)^{\beta+1}\right),
\end{align}
where $Q(a,x)$ is the incomplete gamma function
\begin{align}
  Q(a,x)=\int_x^\infty t^{a-1}e^{-t}dt.
\end{align}

\begin{figure*}
  \centering
  \includegraphics[width=0.8\linewidth]{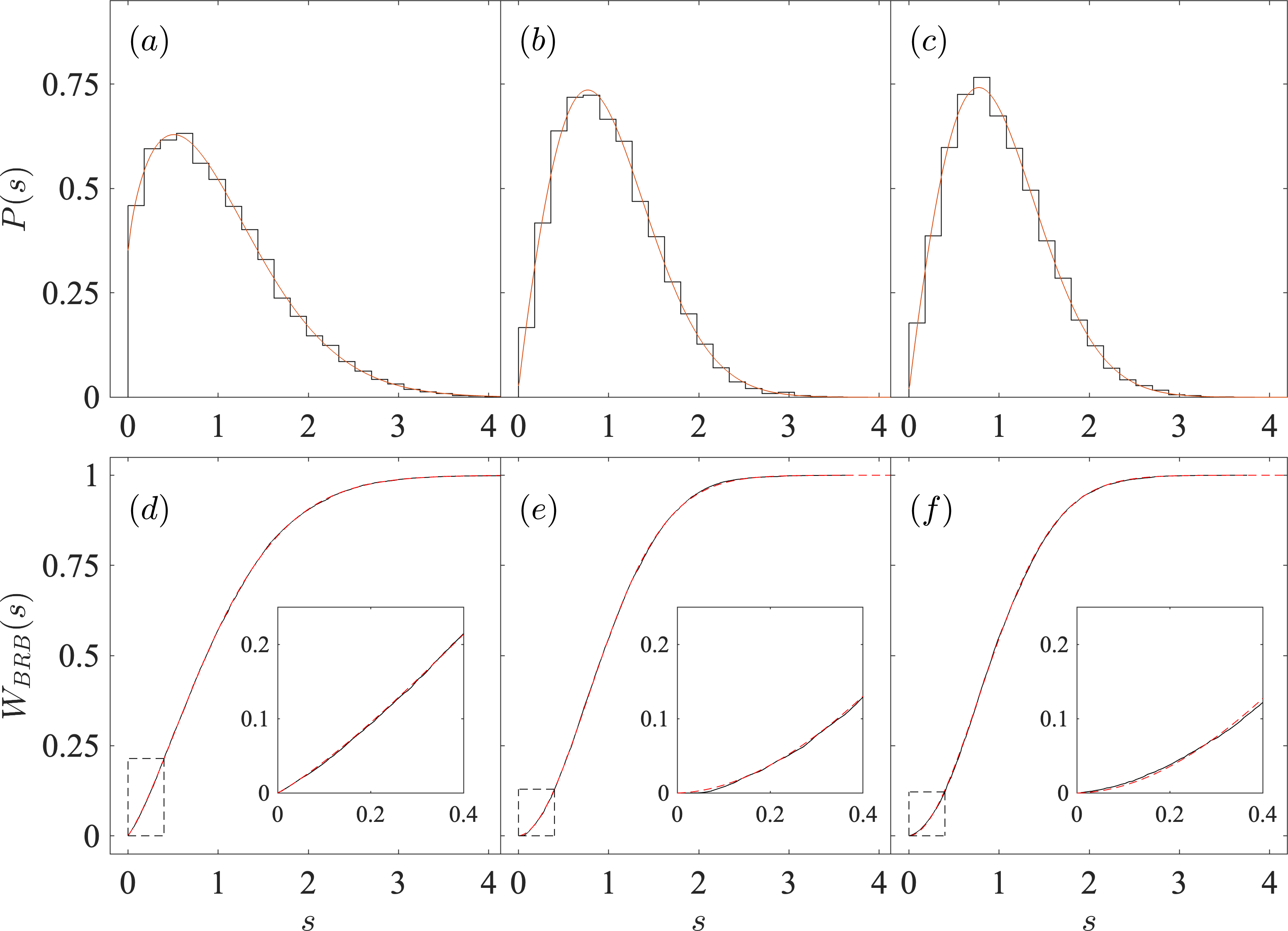}
  \caption{Analogous data as in Fig. \ref{fig:HenonBRB},  but now for the general case with $\lambda=1/16$, from left to right at three energies $E=0.2, 1/3, 0.4$, the width of energy intervals are $\delta E/E\simeq 0.08, 0.03, 0.02$, comprising about 7800 levels each from the doublets.  The fitting parameters $(\beta, \rho_c)$ from left to right are: (0.681, 0.818), (0.926, 0.992), (0.940, 0.995), while the classical fraction $\mu_c$ obtained from SALI is 0.825, 1 and 1, respectively. In both panels, we set the cutoff of $n$ as $N=2200$ and $\hbar=4\times 10^{-4}$, the resulting $\mathcal{N}_D\sim 1.6\times 10^6$.}
  \label{fig:GenBRB}
\end{figure*}

Consider a mixed-type system with only one dominating chaotic region with fraction $\rho_c$, which is the relative density of the chaotic levels. In the sense of the Rosenzweig-Porter approach to obtain the Berry-Robnik distribution, in which the regular and chaotic eigenstates are assumed to be uncorrelated (no tunneling), the gap probability which includes the localization effects is then  written as
\begin{align}
  \mathcal{E}(s)=\mathcal{E}_p(\rho_r s)\mathcal{E}_B(\rho_c s),\quad \rho_r+\rho_c=1,
\end{align}
where $\rho_r$ denotes the relative density of the regular levels, the gap probability for regular spectrum $\mathcal{E}_p(s)= \exp(-s)$. In the semiclassical limit, we expect $\rho_c=\mu_c$ which indeed is confirmed below. The Berry-Robnik-Brody distribution given as the second derivative of the gap probability,
\begin{align}
  P(s)=\frac{d^2\mathcal{E}(s)}{ds^2},
\end{align}
can capture both the relative size of divided quantum phase space, and the localization of the chaotic eigenstates.

The BRB distribution is perfectly confirmed in our numerical calculations, as shown in Figs. \ref{fig:HenonBRB}-\ref{fig:GenBRB} both the levels spacing distributions and the cumulative level spacing distributions, for three energy intervals  $[E-\delta E/2, E+\delta E/2]$, each comprising about 7800 consecutive  levels from the doublets: of $\alpha$-FPUT with $\alpha=1$, $E=0.14,0.15,0.16$, while for the general case $E=0.2, 1/3, 0.4$. In both cases, we have set the cutoff of $n$ and $\hbar$ accordingly, to make sure that the number of energy levels below $E_s$ is less than 1/3 of the available levels for calculation, to achieve good convergence of the numerical spectra. The width of every interval fulfills $\delta E/E \ll 1$, so that all eigenstates from the same interval correspond to almost the same classical dynamics. 

The goodness of the BRB statistics is clearly shown in Figs. \ref{fig:HenonBRB}-\ref{fig:GenBRB}. In the inset, magnifications of cumulative level spacing distributions are shown, which display minor deviations of the numerical data from the best fitting BRB distributions. Our calculation shows that difference between the fitting parameter $\rho_c$ from BRB statistics and the classical fraction $\mu_c$ we have obtained from SALI in Sec. \ref{sec3.3} is less than $1\%$, except for energy $E=0.14$ in $\alpha$-FPUT, where the difference is about $6\%$. This discrepancy would be diminished if we go further deeper into the semiclassical limit. 

Another observation is that the value of $\beta$ increases monotonically with the energy, for both cases: it increases to 0.882 at $E=0.16$ in $\alpha$-FPUT, while it increases to 0.940 in the general case, a value close to 1. This monotonically increasing behavior is predicable, from the estimation of the ratio $t_H/t_T$. The Heisenberg time $t_H$ can be approximated as
\begin{align}
  t_H=2\pi \hbar \rho(E) \simeq 2\pi \hbar g(E),
\end{align}
where $g(E)$ is the smoothed quantum DOS, of which we have derived the analytical expression in Sec. \ref{sec5.1} and shown that the scaled available volume of phase space $2\pi \hbar^2g(E)$ is a monotonically increasing function of $E$.  It follows that $t_H$ is also a monotonically increasing function of $E$. Moreover, the classical fraction $\mu_c$ as an indicator of classical chaos shown in Fig. \ref{fig:fraction} is monotonically increasing with three energies in $\alpha$-FPUT, and not decreasing for three energies in the general case, while the classical transport time $t_T$ is expected to decrease with higher degree of chaos, thus in both cases $t_T$ is at least not decreasing with larger $E$. From the above estimation of $t_H$ and $t_T$, we can conclude that the ratio $t_H/t_T$ increases with larger energy. Due to the direct relation between $t_H/t_T$ and $\beta$, that is, larger value of this ratio results in greater degree of delocalization, i.e. larger value of $\beta$. 

\section{Conclusions and discussion}
\label{sec7}

We have presented an integral view on the transition to chaos of the three-particle FPUT model, for both the $\alpha$-type and the general case, by a  study of emergence of chaos and ergodicity across the energy, classically and quantally. On the classical side, after the introduction of normal modes, we have obtained the relation between $\mu_c$ -- the relative Liouville volume measure of the chaotic part of the phase space -- and the energy $E$, by the SALI method. It shows all classical regimes of interest: the almost entirely regular, the mixed-type regimes, and the entirely chaotic. In the quantum picture, we introduced the rotated bosonic operators, then used the quantum DOS from the numerical quantum spectra obtained by the Krylov subspace method, and the one from quantum typicality, to compare with the analytical results of smoothed DOS from the Thomas-Fermi rule. 

The derivatives of the semiclassical DOS show two types of singularity, also referred to as the ESQPT. While the logarithmic divergence results from the saddles (unstable fixed points) of the classical Hamiltonian, the step discontinuity corresponds directly to the stable fixed point, and both can be clearly understood from the geometric change of the available phase space.

Based on the nice agreement between the numerical DOS and the semiclassical results, the nearest-level spacing distribution (of the doublets from $C_{3v}$ symmetry) after unfolding, as well as the mean spacing ratio, show a transition from the Poisson to the Wigner-Dyson across the energy spectrum. Moreover, the normalized mean spacing ratio demonstrates a functional dependency on the classical fraction $\mu_c$. Further in deeper semiclassical limit, for the first time, we have verified the BRB statistics in a continuous Hamiltonian system, where the extracted quantum Berry-Robnik parameter is found to agree with the classical value within better than one percent. The Brody parameter in all cases, especially in the $\alpha$-type, is not exactly close to 1, as a consequence of dynamical phase space localization, which we will study in more detail by using different types of Husimi function representations in paper II. 

This work in a way has provided the energy-resolved correlation between the classical route to chaos and the spectral statistics of quantum spectra. We noticed that a recent study on Bose-Hubbard Hamiltonian \cite{pausch2021chaos} has provided the energy-resolved correlation between the spectral features and structural changes of the associated eigenstates measured by generalized fractal dimensions (GFD) defined in Fock space. It is an interesting topic for the future work to explore the difference between structural changes measured by GFD defined in two-mode circular (Fock) basis and by localization measures defined in phase space, which have a clear correspondence with the classical dynamics.

One technical issue needs to be addressed. In the calculation of quantum spectra in narrow energy intervals across the energy spectrum, we have employed the Krylov subspace method for large sparse matrices, a shift-invert method of exact diagonalization, for which there would be a significant drawback (larger memory requirements) at larger cutoff $N$. The polynomially filtered exact diagonalization approach \cite{sierant2020polynomially} is proved to scale better with system size, which is an interesting task for a future research to employ this approach for the study of spectral statistics and entanglement entropy in deeper semiclassical limit.

\section{Acknowledgement}
We thank  H. Skokos for useful discussion of the SALI method. This work was supported by the Slovenian Research and Innovation Agency (ARIS) under the grant J1-4387.

  \appendix
  \section{The circular two-mode basis}
  \label{appA}
  It can be verified that there are commutation relations
  \begin{equation}
    \begin{aligned}
      [\hat{\ell},a_+]&=-\hbar a_+,\quad [\hat{\ell},a_-]=\hbar a_-, \\ 
       [\hat{n},a_+]&=-\hbar a_+,\quad [\hat{n},a_-]=-\hbar a_-.
  \end{aligned}
  \end{equation}
With  $[\hat{n},\hat{\ell}]=0$, $\hat{n}$ and $\hat{\ell}$ possess simultaneous eigenfunctions $ |n,l\rangle$ with real eigenvalues
\begin{align}
    \hat{n} |n,l\rangle=n\hbar|n,l\rangle, \quad \hat{\ell}|n,l\rangle=l\hbar|n,l\rangle,
\end{align}
and for rotated bosonic operators
\begin{equation}
  \begin{aligned}
    \hat{\ell} a_+|n,l\rangle&=\hbar(l-1)a_+ |n,l\rangle,  \\
    \hat{n}a_+|n,l\rangle&=\hbar(n-1)a_+|n,l\rangle.
\end{aligned}
\end{equation}
Consequently,
\begin{align}
    a_+|n,l\rangle&=A_{nl}|n-1,l-1\rangle,\ a_-|n,l\rangle=B_{nl}|n-1,l+1\rangle, \nonumber \\
    a_+^\dagger|n,l\rangle &=C_{nl}|n+1,l+1\rangle,\ a_-^\dagger|n,l\rangle = D_{nl}|n+1,l-1\rangle, \nonumber
\end{align}
where $A_{nl}$ and $B_{nl}$ are normalizing factors depending on $n$ and $l$, and it can be easily verified $A_{nl}=C_{n-1,l-1}^*, B_{nl}=D_{n-1,l+1}^*$, with $(a_-a_+^\dagger)^k|n,l\rangle$ and $(a_+a_-^\dagger)^k|n,l\rangle\ (k\ge 1)$ being the  eigenfunctions of $\hat{n}$ and $\hat{\ell}$, belonging to the same energy $n$, but to different $\hat{\ell}- \text{eigenvalues}$ $(l+2)\hbar,(l+4)\hbar,\dots, (l-2)\hbar,(l-4)\hbar,\dots$ Further analysis of the upper and lower bound of $l$ and $n$ shows that \cite{louck1960generalized}
\begin{align}
    \hat{n}|n,l\rangle = (n+1)\hbar|n,l\rangle , \quad \hat{\ell}|n,l\rangle=l\hbar|n,l\rangle,  
\end{align}
with $l=n,n-2,\dots,-n \ (n\in \mathbb{N}_0)$ and
\begin{align}
    \label{eq:quantum-phase}
    A_{nl}=e^{i\delta}\sqrt{(n+l)/2}, \quad B_{nl}=e^{i\gamma}\sqrt{(n-l)/2},
\end{align}
where we choose $e^{i\delta}=e^{i\gamma}=1$ ($\delta$ and $\gamma$ are arbitrary, choosing different $\delta$ and $\gamma$ does not change the quantum spectra), with this choice of phase factors, 
\begin{widetext}
\begin{equation}
  \begin{aligned}
    \label{eq:quantum-coeff1}
    a_+|n,l\rangle &=\sqrt{(n+l)/2}|n-1,l-1\rangle,\quad  a_-|n,l\rangle =\sqrt{(n-l)/2}|n-1,l+1\rangle,\\
    a_+^\dagger |n,l\rangle &=\sqrt{(n+l+2)/2}|n+1,l+1\rangle,\quad  a_-^\dagger|n,l\rangle =\sqrt{(n-l+2)/2}|n+1,l-1\rangle. 
\end{aligned}
\end{equation}
Using these expressions of (rotated) annihilation and creation operator, according to Eq. \eqref{eq:rotated-bosonic}, we get 
\begin{equation}
  \begin{aligned}
    \label{eq:quantum-coeff2}
    \hat{q}_+|n,l\rangle &=\sqrt{\hbar}(a_+^++a_-)|n,l\rangle =\sqrt{\hbar}[\sqrt{(n+l+2)/2}|n+1,l+1\rangle+\sqrt{(n-l)/2}|n-1,l+1\rangle],\\
    \hat{q}_-|n,l\rangle &=\sqrt{\hbar}(a_++a_-^\dagger)|n,l\rangle =\sqrt{\hbar}[\sqrt{(n+l)/2}|n-1,l-1\rangle+\sqrt{(n-l+2)/2}|n+1,l-1\rangle ].
\end{aligned}
\end{equation}
\end{widetext}

\section{Representation of states and symmetries}
\label{appB}
Classical Hamiltonian system of the three-particle FPUT is of $C_{3v}$ symmetry and is time-reversal invariant. In the quantum Hamiltonian, the time-reversal operator in the circular-mode basis acts as
\begin{align}
    T|n,l\rangle =|n,-l\rangle.
\end{align}
It is easy to verify that $T a_\pm^\dagger T^{-1}=a_\mp^\dagger$, $T a_\pm T^{-1}= a_\mp$, from basis physical quantities: $TqT^{-1}=q$, $TpT^{-1}=-p$, $TiT^{-1}=-i$. One can define the basis states
\begin{align}
|n,l,s\rangle=a_{l,s}(|n,l\rangle +s|n,-l\rangle)/\sqrt{2}, \quad s=\pm 1,
\end{align}
where $a_{l,s}=s^{\text{mod}(l,3)}$. Two properties of this basis should be noted: (i) there are no states with $l=0$ and $s=-1$. (ii) The states $|n,l,s\rangle$ and $|n,-l,s\rangle$ are linearly dependent, with the relation $|n,-l,s\rangle =a_{l+1,s}|n,l,s\rangle$ (proved by the equality $a_{l+1,s}=sa_{-l,s}/a_{l,s}$), thus the basis $\{|n,l,s\rangle, l\ge 0\}$ can be used as a complete set, and $\langle n',l',s'|n,l.s\rangle =2^{\delta_{l,0}}\delta_{n,n'}\delta_{l,l'}\delta_{s,s'}$.  To establish this basis as orthonormal, we make the rescaling $a_{l,s}=2^{-\delta_{l,0}/2}s^{\text{mod}(l,3)}$. These basis states are also the eigenstates of $T$
\begin{align}
    T|n,l,s\rangle =\frac{a_{l,s}}{\sqrt{2}}(T|n,l\rangle +sT|n,-l\rangle)=s|n,l,s\rangle.
\end{align}
 The operator $ P_s =\frac{1}{2}(1+sT)$
is a projection operator to the subspace spanned by the states $\{|n,l,s\rangle\}$, and $|n,l,s\rangle =\sqrt{2}a_{l,s}P_s|n,l\rangle$. From the relation $T\hat{q}_+T^{-1}=\hat{q}_-$, it is straightforward to verify that
\begin{align}
\hat{q}_+^3|n,l\rangle=T\hat{q}_-^3T^{-1}|n,l\rangle=T\hat{q}_-^3|n,-l\rangle,
\end{align}
which has also been proven in Eq. \eqref{eq:coeff-cubic}. 
It is then evident that
  \begin{align}
    [T,i(\hat{q}_-^3-\hat{q}_+^3)] =0,\ 
    [P_s,i(\hat{q}_-^3-\hat{q}_+^3)]=0.
\end{align}
For simplicity, here we set the phase in Eq. \eqref{eq:quantum-phase} as $\delta=\pi/2, \gamma=-\pi/2$, then
\begin{align}
    (i\hat{q}_+)^3|n,l\rangle =\hbar^{3/2}\sum_{m=\pm1,\pm3} k_m^+ (n,l)|n+m,l+3\rangle,
\end{align}
so finally, using the equality $a_{l,s}=a_{l\pm3, s}$ and two properties of the basis,  for $l\ge 0$ we get
\begin{align}
    i(&\hat{q}_-^3-\hat{q}_+^3)|n,l,s\rangle =\sqrt{2}a_{l,s}\left[(i\hat{q}_+)^3+(-i\hat{q}_-)^3\right]P_s|n,l\rangle\nonumber \\
    &=\sqrt{2}a_{l,s}P_s\left[(i\hat{q}_+)^3+(-i\hat{q}_-)^3\right]|n,l\rangle\nonumber\\
    &=\hbar^{3/2}\sum_{m=\pm1,\pm3} k_m^\pm (n,l)|n+m,l\pm3,s\rangle\nonumber \\&=\hbar^{3/2}\sum_{m=\pm1,\pm3} k_m^\pm (n,l)|n+m,|l\pm3|,s\rangle.
\end{align}
 It reveals that the basis states $|n,l,s\rangle$ with different $s$ are decoupled. The circular basis states are coupled if and only if $\Delta l=\pm 3$: the states with $\text{mod}(l,3)=0$ are decoupled from the states with $\text{mod}(l,3)\ne 0$. 

\section{Available phase space for the three-particle FPUT}
\label{appC}
The Thomas-Fermi expression for DOS of the three-particle generic FPUT is given as
\begin{align}
  &g(E)=\frac{1}{(2\pi\hbar)^2}\int d\textbf{q}d\textbf{p}\delta(E-H(\textbf{q},\textbf{p}))\nonumber \\
  &=\frac{1}{(2\pi\hbar)^2}\int d\textbf{q}d\textbf{p}\left[\frac
      {\delta (p_1-p_1^+)}{|\partial H/\partial p_1|_{p_1^+}} + \frac
      {\delta (p_1-p_1^-)}{|\partial H/\partial p_1|_{p_1^-}} \right],\nonumber
\end{align}
where $p_1^\pm$ are two roots of the equation $H(\textbf{q},\textbf{p})=E$,
\begin{align}
  p_1^\pm = \pm \sqrt{2E-2V(\textbf{q},\textbf{p})-p_2^2)},
\end{align}
and $|\partial H_\alpha/\partial p_1|_{p_1^\pm}=|p_1^\pm|$, then the $p_1$ integration yields
\begin{align}
  \label{eq:TF-smooth}
  g(E)&=\frac{2}{(2\pi\hbar)^2}\int dq_1dq_2\int_{p_2^-}^{p_2^+}dp_2 \frac{1}{\sqrt{(p_2^+-p_2)(p_2-p_2^-)}}\nonumber\\&=\frac{2\pi}{(2\pi\hbar)^2}\int dq_1dq_2=\frac{1}{2\pi\hbar^2}\int r drd\phi,
\end{align}
with $p_2^\pm=\pm\sqrt{2E-2V(\textbf{q},\textbf{p})}$, and the constraint of the integral is given as $V(r,\phi)\le E$.  A further simplification of the integral rests upon the derivation of the real roots $r(\phi)$ for $V(r,\phi)=E$. 

For $\alpha$-FPUT of cubic potential, the condition for the existence of real roots ($0\le r<\infty$) for the cubic equation 
\begin{align}
\frac{\alpha \sin3\phi }{3}r^3+  \frac{1}{2}r^2 -E=0
\end{align}
is that the discriminant 
\begin{align}
  D_\alpha = \frac{9E}{(2\alpha\sin 3\phi)^2}(E-\frac{1}{6\alpha^2\sin^2 3\phi}) \le 0, 
\end{align}
that is, the energy $E$ is bounded by $E\le 1/6\alpha^2$. The positive real root is
\begin{align}
  \begin{cases}
  r_+(\phi) = \frac{1}{\alpha\sin 3\phi}(\cos\frac{\theta}{3}-\frac{1}{2}), &\text{if } \sin3\phi \ge0 \\
  r_-(\phi)=\frac{-1}{\alpha\sin 3\phi}(\cos\frac{\theta+\pi}{3}+\frac{1}{2}), &\text{if }\sin3\phi \le0,
  \end{cases}
\end{align}
where 
\begin{align}
  \cos\theta = 12E\alpha^2\sin^23\phi-1, \quad \theta\in[0,\pi].
\end{align}
Inserting this result into the simplified Thomas-Fermi expression in Eq. \eqref{eq:TF-smooth}, we get
\begin{align}
  g(E) =\frac{3}{4\pi\hbar^2}\big(\int_0^{\pi/3}r_+^2(\phi)d\phi+\int^{2\pi/3}_{\pi/3}r_-^2(\phi)d\phi\big),
\end{align}

From the evaluation of $g(E)$ we can calculate the derivative $g^\prime(E)$ and obtain
\begin{align}
  g^\prime(E)=3\int_0^{\frac{\pi}{3}}\frac{\sin\frac{2\theta}{3}-\sin\frac{\theta}{3}+\sin\frac{2\theta+2\pi}{3}+\sin\frac{\theta+\pi}{3}}{\pi\hbar^2\sqrt{1-\cos^2\theta}}d\phi.
\end{align}
When $E\to 1/6\alpha^2$, the integral would diverge, because of the singular point $(\phi,\theta)=(\pi/6,0)$. For energy $E=1/6\alpha^2-\delta$ with small $\delta$
\begin{align}
  g^\prime(\frac{1}{6\alpha^2}-\delta) 
  &=\frac{\sqrt{2}}{\pi\hbar^2}\int^\pi_{\theta_\delta}\frac{\sin(\frac{\theta}{6}+\frac{\pi}{3})}{\sqrt{2\mu-1-\cos\theta}}d\theta\nonumber\\
  &=\frac{\sqrt{2}c}{\pi\hbar^2}\int^\pi_{\theta_\delta}\frac{d\theta}{\sqrt{2\mu-1-\cos\theta}}
\end{align}
 where $c$ is a constant that fulfills $\sqrt{3}/2<c<1$, obtained by referring to the mean value theorem in calculating this integral, and we denote
$\mu=1-6\delta\alpha^2$, $\cos\theta=2\mu\sin^23\phi-1$ and $\cos\theta_\delta=2\mu-1$.  The integral 
\begin{align}
  \int^\pi_{\theta_\delta}\frac{d\theta}{\sqrt{2\mu-1-\cos\theta}} =\int_0^{1}\frac{\sqrt{\mu}dt}{\sqrt{1-\mu t^2}\sqrt{1-t^2}},
\end{align}
is Legendre's incomplete elliptic integral of the first kind, with $\delta\to 0$, $\mu\to 1$ the asymptotic approximation gives
\begin{align}
  \int^\pi_{\theta_\delta}\frac{d\theta}{\sqrt{2\mu-1-\cos\theta}} \approx  \frac{\sqrt{\mu}}{2} \ln \frac{16}{1-\mu}.
\end{align}
From exact expression for $g^\prime(E)$ we therefore explicitly extract the logarithmic divergence 
\begin{align}
  g^\prime(\frac{1}{6\alpha^2}-\delta) \approx -\frac{c}{\sqrt{2}\pi\hbar^2}\ln \frac{3\delta \alpha^2}{8}.
\end{align}

For the general case of quartic potential where
\begin{align}
  \label{eq:quartic}
  V(r,\phi)=\frac{1}{2}r^2+\frac{1}{3}r^3\sin3\phi +\lambda r^4 = E,
\end{align}
a quartic equation can be solved in general, thanks to the genius of Abel and Galois. We see from the potential landscape that for $\lambda\ge 1/16$,  there is one positive root $r>0$. It can also be verified from the discriminant 
\begin{align}
    D= &-256 E^3 \lambda ^3-32 E^2 \lambda ^2+8 E^2 \lambda  \sin^2 3 \phi -E\lambda\nonumber\\
    &-\frac{1}{3} E^2 \sin^4 3 \phi +\frac{1}{18} E \sin ^23 \phi < 0,
\end{align}
and Eq. \eqref{eq:quartic} in a depressed cubic form
\begin{equation}
  \begin{aligned}
    t_0^3+pt_0+q=0, \quad p=\frac{4 E}{\lambda }-\frac{1}{12 \lambda ^2}, \\
     q=\frac{E \sin^2 3 \phi}{9 \lambda ^3}-\frac{1}{108 \lambda ^3}-\frac{4 E}{3 \lambda ^2},
\end{aligned}
\end{equation}
where the discriminant scaled as
\begin{align}
    D_3=4p^3+27q^2=-D/\lambda^6>0.
\end{align}
The only real root of the depressed cubic equation then could be expressed by hyperbolic functions as
\begin{align}
    \label{eq:append-hyperbolic}
    t_0&=-\frac{2|q|}{q}\sqrt{\frac{-p}{3}}\cosh\big[\frac{1}{3}\arccosh(\frac{-3|q|}{2p}\sqrt{\frac{-3}{p}})\big], p<0 ,\nonumber\\
    t_0&=-2\sqrt{\frac{p}{3}}\sinh\big[\frac{1}{3}\arcsinh(\frac{3q}{2p}\sqrt{\frac{3}{p}})\big], \text{if} \ p>0.
\end{align}
We denote 
\begin{align}
    p_0=\frac{1}{2 \lambda }-\frac{\sin^2 3 \phi}{24 \lambda ^2},\quad q_0=\frac{\sin^3 3 \phi }{216 \lambda ^3}-\frac{\sin 3 \phi }{12 \lambda ^2},
\end{align}
and $t=t_0-2p_0/3$, the root reads
\begin{align}
    \begin{cases}
    r_+(\phi) = \frac{\sqrt{t}+
    \sqrt{t-2(p_0+t+q_0/\sqrt{t})}
    }{2}-\frac{\sin3\phi}{12\lambda}, \ &\sin3\phi \ge0, \\
    r_-(\phi)=\frac{-\sqrt{t}+
    \sqrt{t-2(p_0+t-q_0/\sqrt{t})}}{2}-\frac{\sin3\phi}{12\lambda}, \ &\sin3\phi \le0. 
    \end{cases} 
    \label{eq:roots-pm}
\end{align}
the Thomas-Fermi expression is then given as
\begin{align}
    \label{eq:app2-TF}
    g(E) =\frac{3}{4\pi\hbar^2}\int_0^{\frac{\pi}{3}}\big( \frac{\sin^2 3 \phi}{18 \lambda ^2}-\frac{1}{2\lambda}-\frac{\sin 3 \phi}{6 \lambda }\sqrt{t}-\frac{q_0}{\sqrt{t}}\big)d\phi,
\end{align}
with the derivative 
\begin{align}
    g^\prime(E)=-\frac{3\pi}{(2\pi\hbar)^2}\int^{\pi/3}_0\partial_E(\frac{\sin 3 \phi}{6 \lambda }\sqrt{t}+\frac{q_0}{\sqrt{t}})d\phi.
\end{align}

For $\lambda<1/16$, there would exist multiple roots below the saddle energy. From an algebraic perspective, in this scenario, there are sets of $\phi$ for which the discriminant of the quartic equation $D>0$, for specific value of $E$. Upon deeper examination of the potential landscapes, it becomes evident that the parameter space of $\lambda$ can be partitioned into two distinct regions: $0<\lambda<1/18$ and $1/18<\lambda<1/16$. In the first region, as the energy increases, a notable transformation occurs in the geometry of the available phase space: it evolves from three disconnected parts to four disconnected parts, ultimately culminating in an interconnected configuration. In the subsequent region, there is a transition from a single connected part to four disconnected parts, before eventually returning to the interconnected. The global minimum of energy in this scenario is the minimum energy among all stable fixed points 
$ E_{min}=\min\{V(\lambda_+,\phi_m), 0\}$.
The nature of roots of Eq. \eqref{eq:quartic} varies across energies defined as the following
\begin{align}
  E_l&= \max\{V(\lambda_+,\phi_m),0\},\ E_s=V(\lambda_-,\phi_m), \nonumber\\
     E_1&=V(1/\sqrt{4\lambda},\phi)|_{\sin3\phi=-\sqrt{16\lambda}},
\end{align}
where $E_l >0$  is the (local minimum) energy at the stable fixed points, $E_s$ denotes energy of saddles, and the discriminant $D< 0$ when $E\ge E_1$.

 In the first region $\lambda<1/18$, there are two roots for the classical available phase space composed of three disconnected parts for $E_{min}<E<E_l$
 \begin{equation}
    r_{u,d} = \frac{\sqrt{t}\pm
    \sqrt{t-2\left(p_0+t+q_0/\sqrt{t}\right)}}{2}-\frac{\sin3\phi}{12\lambda},
\end{equation}
where $-1\le\sin3\phi<\mu$, the sign $+$ is for $r_u$ and the minus for $r_d$,  $\mu$ is the solution of the following equation 
\begin{align}
    V(\frac{-\sin3\phi+\sqrt{\sin^23\phi-16\lambda}}{8\lambda})|_{\sin3\phi=\mu}=E.
\end{align}
So finally, one has the density of states as an integral 
\begin{align}
    g(E)=\frac{3\pi}{2\pi\hbar^2}\int_{\frac{1}{3}(\arcsin\mu-\pi)}^{\pi/2} (r_u^2-r_d^2) d\phi.
\end{align}
For $E_l<E<E_s$, there are three real positive roots: $r_{u,d}(\phi)$ for the outer three disconnected parts with
\begin{align}
    t_0=2\sqrt{\frac{-p}{3}}\cos\big[\frac{1}{3}\arccos(\frac{3q}{2p}\sqrt{\frac{-3}{p}})\big], 
\end{align}
and $r_i(\phi)$  for the inner connected part with
\begin{align}
t_0=2\sqrt{\frac{-p}{3}}\cos\big[\frac{1}{3}\arccos(\frac{3q}{2p}\sqrt{\frac{-3}{p}})+\frac{2\pi}{3}\big],
\end{align}
of the same form as Eq. \eqref{eq:roots-pm}, depending on the sign of  $\sin3\phi$. The density of states  is
\begin{align}
    g(E)=\frac{1}{4\pi\hbar^2}\big(6\int_{\frac{1}{3}(\arcsin\mu-\pi)}^{\pi/2} (r_u^2-r_d^2) d\phi +\int_0^{2\pi} r_i^2d\phi\big).
\end{align}
For $E_s<E<E_1$, the available phase space is a connected one piece, yielding
\begin{align}
    &g(E)=\frac{3}{2\pi\hbar^2}\int_{\frac{1}{3}(\arcsin\mu-\pi)}^{\frac{1}{3}(\arcsin\nu-\pi)} (r_u^2-r_d^2) d\phi \nonumber\\
    &\ +\frac{1}{4\pi\hbar^2}\big(3\int_{\frac{1}{3}(\arcsin\nu-\pi)}^{\frac{1}{3}(\arcsin\nu+2\pi)}r_u^2 d\phi
    +\int_0^{2\pi} r_i^2d\phi\big),
\end{align}
where $\nu$ is the solution of the following equation 
\begin{align}
    V(\frac{-\sin3\phi-\sqrt{\sin^23\phi-16\lambda}}{8\lambda},\phi)|_{\sin3\phi=\nu}=E. 
\end{align}
As energy goes beyond $E_1$, the roots are given again as Eq. \eqref{eq:roots-pm} where the discriminant $D<0$. By applying a similar approach, we can have an analytical expression for the region in the interval $1/18<\lambda<1/16$.

  \bibliographystyle{apsrev4-2}
  \bibliography{yr.bib}

  \end{document}